\UseRawInputEncoding
\documentclass{article}
\usepackage{graphicx} 

\usepackage{amsmath}
\usepackage{multirow}
\usepackage{tikz}
\usepackage{graphicx}
\usepackage{subcaption}  
\usepackage[margin=1in]{geometry}
\usepackage{amsmath}
\usepackage{listings}
\usepackage{booktabs} 
\usepackage{booktabs}
\usepackage{multirow}
\usepackage{algorithm}
\usepackage[noend]{algpseudocode}
\usepackage[backend=biber, sorting=none]{biblatex}
\author{Lopamudra Dey}

\usepackage{authblk}

\title{WTKO-CNN: Deep Learning Reveals Sequence Motifs Distinguishing Wild-Type and Knockout ATAC-seq Peaks}
\date{}
\author[1]{Lopamudra Dey}

\affil[1]{Department of Biomedical and Clinical Sciences, Division of Molecular Medicine and Virology (MMV), Linkoping University, Sweden, Email: lopamudra.dey1@gmail.com}

\begin{document}

\maketitle

\section{Abstract}
Chromatin accessibility is a key determinant of transcriptional regulation, and perturbation of chromatin regulators can lead to distinct regulatory landscapes between wild-type (WT) and knockout (KO) conditions. Identifying the sequence features that drive these condition-specific differences remains a major challenge. Here, I develop \textbf{WTKO-CNN}, an interpretable deep learning framework based on a one-dimensional convolutional neural network with an attention mechanism to classify regulatory DNA sequences as WT- or KO-specific from ATAC-seq datasets. The model was evaluated across three chromatin accessibility datasets associated with RelA, LSD1, and Ebf1 perturbations, demonstrating its ability to distinguish WT and KO regulatory states. To understand the sequence features learned by the model, I applied attention-guided saliency analysis to identify nucleotide regions contributing most strongly to classification decisions. High-contribution sequence regions were extracted as k-mers from CNN saliency and filters, clustered, and analyzed using de novo motif discovery approaches. The resulting motifs were characterized using MEME, TOMTOM, and HOMER to identify similarities with known transcription factor binding profiles. CNN-derived motifs revealed biologically relevant sequence patterns associated with transcription factor families involved in transcriptional regulation and chromatin accessibility, including AP-1, SP, KLF, Runt, and EBF-associated motifs. Together, these results demonstrate that WTKO-CNN provides both predictive performance and mechanistic interpretability, enabling the discovery of sequence determinants underlying WT and KO-specific chromatin accessibility. This framework offers a complementary approach to conventional motif enrichment methods by uncovering both canonical and non-canonical regulatory elements that may contribute to transcriptional changes following chromatin perturbation.


\section{Introduction}
The Transposase Accessible Chromatin Assay (ATAC-seq) data is the most recent significant method to assess genome-wide chromatin accessibility accurately. Over the years, both determining areas that are differentially accessible (DA) between two cellular states and identifying basal accessible chromatin regions within a specific cellular context have been accomplished using ATAC-seq \cite{5}. Transcription factors (TF), which bind to particular DNA sequences to modify transcriptional activity, are crucial for the control of genes \cite{4}. Understanding the relationship between transcription factor (TF) binding and chromatin accessibility is essential for understanding transcriptional regulation, cell state control, and the emergence of novel phenotypes. TFs can identify their motifs and control gene expression programs in putatively open regions that have been cataloged by recent genome-wide chromatin accessibility profiling studies \cite{2,3,6}. 

Deep learning models have been increasingly popular in recent years and have produced ground-breaking findings in a wide range of scientific fields \cite{9,10,11,12,32, 33}. Due to their intricate architecture, these deep learning models are frequently referred to as "black boxes" despite their excellent predictive ability \cite{34}. This "black box" style makes it challenging to understand how predictions are made. However, model interpretability is critical in biological research since prediction accuracy is just as important as comprehending the underlying sequence characteristics and regulatory logic controlling gene expression. To open this “black box,” I propose \textbf{WTKO-CNN}, a two-layer convolutional neural network with a saliency map that exploits k-mer representations derived from ATAC-seq data to reveal sequence determinants underlying differences between wild-type (WT) and knockout (KO) chromatin accessibility regions. By facilitating the direct extraction of informative k-mer patterns, WTKO-CNN prioritises interpretability over traditional deep learning approaches. The learned convolutional filters correspond to sequence motifs that provide mechanistic insight into regulatory alterations induced by gene knockout.


Recently, deep learning approaches, particularly convolutional neural networks (CNNs) have been applied to genomic sequences, enabling automated feature extraction and the identification of sequence regions most predictive of TF binding through saliency maps \cite{29}. These methods are used in many research works for better peak detection and classification compared to traditional peak-calling tools such as MACS2 \cite{8,9}. In many research works, it can be seen that combining CNN and LSTM in a hybrid model is a powerful approach for DNA sequence classification tasks \cite{11,12}. In \cite{13}, the authors suggested deep learning techniques such as CNN, DNN, and the N-gram probabilistic model to classify DNA sequences. Based on the distance measure, a novel method for feature extraction using random DNA sequences put forth. The CNN layers are responsible for extracting local sequence features, such as short k-mers or motifs, by applying convolutional filters to the one-hot encoded DNA sequences. These local patterns are then downsampled using pooling layers to reduce dimensionality \cite{14}. In \cite{27} the authors introduced DeepSEA, a deep convolutional neural network capable of predicting chromatin effects directly from DNA sequence. The model demonstrated that deep learning can automatically learn regulatory sequence features without manual feature engineering. Later, interpretability became an important research direction. In \cite{19}, an interpretability model is provided by the Deep Motif Dashboard (DeMo Dashboard), which aims to clarify how deep neural networks categorise transcription factor binding sites (TFBS). They evaluated three methods: convolutional (CNN), recurrent (RNN), and hybrid CNN-RNN architectures and, using saliency maps, they identified high-contribution nucleotides via first-order derivatives, and emporal output scores. Zhang et al. proposed a new architecture CAE-CNN (Convolutional AutoEncoder and Convolutional Neural Network) by combining a convolutional autoencoder with convoluational neural network. They used a highway connection layer to capture DNA nucleotide patterns, and extensive evaluations on human and mouse TFBS datasets demonstrate that an integrated unsupervised-supervised approach outperforms state-of-the-art methods in motif discovery with accuracy, precision, recall, and AUC \cite{18}. In \cite{20}, the authors showed that motif extraction from deep CNNs is difficult due to its scattered representations and also improved test performance does not ensure interpretability. They tested CNNs on synthetic and actual DNA sequences and demonstrates that adding exponential activations to first-layer filters results in more robust and interpretable motif representations. DeepSite uses CNN and bidirectional long short-term memory to identify long-term relationships between DNA sequence motifs. Apart from taking sequence dependencies into account, they found how to get motif information in the imbalance data and filter out legitimate information in huge data \cite{22}. To maximise motif prediction, Yang et al. used binomial distribution and deep neural networks in \cite{23}. There are certain issues with prediction accuracy and runtime performance, despite the fact that current deep learning techniques have produced good results in TFBS discovery. In \cite{24} the authors provide a novel two-step model called the Salient Relevance (SR) map, which attempts to clarify how deep CNNs identify images and extract characteristics from regions known as attention areas. Another work, DeepLIFT (Deep Learning Important FeaTures) decomposed the contributions of individual nucleotides within one-hot encoded sequences to the final model output. The model explains neural network predictions by tracing how differences from a baseline input propagate through the network and assigning importance scores to each input feature \cite{25}. To improve transparency, \cite{26} proposed ExplaiNN, an interpretable neural network architecture for genomics. ExplaiNN consists of multiple independent convolutional units, where each unit learns a motif-like feature from DNA sequences. These features are combined through a linear layer, allowing researchers to easily interpret the contribution of each learned motif to the final prediction. Other model explanation techniques, such as SHAP and Integrated Gradients, have also been widely used to interpret neural network predictions in genomics. These approaches help identify biologically meaningful sequence motifs and regulatory patterns. Overall, recent research focuses on building models that not only achieve high predictive performance but also provide biological interpretability \cite{35}. In \cite{37} TAMC employs a deep-learning approach to predict motif-centric TF binding activity using paired-end ATAC-seq data. It does not need bias correction during signal processing. Using a one-dimensional convolutional neural network (1D-CNN), it makes predictions based on footprint and non-footprint features at binding sites for each TF. It outperforms existing footprinting tools in TFBS prediction, especially for ATAC-seq data with low sequencing depth.

 Recurrent neural networks (RNNs) are capable of modeling sequential dependencies in DNA sequences; however, their interpretability remains limited, and they are less effective for identifying short, position-independent sequence motifs that often characterize transcription factor binding sites. Convolutional neural networks (CNNs), in contrast, are well suited for regulatory genomics because their convolutional filters naturally capture local k-mer-like sequence patterns associated with transcription factor binding and chromatin regulatory elements. Although hybrid CNN-LSTM architectures can model higher-order dependencies between sequence motifs, they often introduce additional complexity without providing clear improvements in biological interpretability. Therefore, I adopted a CNN-based framework to prioritize interpretable sequence feature discovery.

Despite the predictive power of CNN models, understanding the biological features driving model decisions remains a major challenge. Interpreting learned sequence representations is essential for connecting computational predictions with underlying regulatory mechanisms. Saliency-based approaches provide nucleotide-level resolution by identifying sequence positions that contribute most strongly to model predictions, enabling the discovery of discriminative sequence regions associated with regulatory changes. Complementarily, analysis of CNN filter activations reveals recurrent sequence patterns learned across datasets, allowing the identification of regulatory motifs and potential transcription factor binding signatures. Together, these approaches provide a comprehensive strategy for linking deep learning predictions with interpretable sequence features.

The model, WTKO-CNN, an interpretable CNN-based framework can identify sequence determinants that distinguish WT vs KO regulatory landscapes. Unlike approaches that focus solely on predictive performance, WTKO-CNN integrates saliency-guided k-mer extraction, sequence clustering, and CNN filter-based motif discovery to identify discriminative motifs associated with differential chromatin accessibility and transcriptional regulation. This combined strategy enables systematic characterization of regulatory elements underlying WT and KO differences and provides insights into transcription factor networks that contribute to altered genomic regulation.

\section{Materials and Methods}

\subsection{Dataset}
I evaluated the performance of our convolutional neural network (CNN) using two distinct biological paradigms of transcriptional and epigenetic regulation. These datasets are retrieved from the Gene Expression Omnibus (GEO) and consist of differentially accessible regions (ATAC-seq)  comparing WT and KO in \textit{Mus musculus}. 

\begin{itemize}
    \item \textbf{GSE107075 (RelA):} ATAC-seq data for RelA knockout (KO) and wild-type (WT) mouse fibroblasts were obtained from the Gene Expression Omnibus (GEO) under accession number GSE107075 \cite{17} (\url{https://www.ncbi.nlm.nih.gov/gds/?term=GSE107075}). The dataset was generated using the Illumina NextSeq 500 sequencing platform and contains genome-wide chromatin accessibility profiles from WT and RelA-KO fibroblasts, with two biological replicates per condition. The study investigated changes in chromatin accessibility associated with the deletion of RelA, a key component of the NF-$\kappa$B signaling complex. Accessible chromatin regions identified from ATAC-seq peaks were used to extract genomic sequences for downstream analysis and CNN-based prediction of WT- and KO-specific regulatory patterns.

    \item \textbf{GSE119222 (LSD1):} ATAC-seq data from germinal center (GC) B cells in \textit{Mus musculus} were obtained from GEO accession GSE119222 \cite{28} (\url{https://www.ncbi.nlm.nih.gov/geo/query/acc.cgi?acc=GSE119222}). The dataset includes chromatin accessibility profiles from two WT and three LSD1 knockout samples generated from sorted GC B splenocytes collected ten days after immunization. These data were used to characterize accessibility changes associated with loss of the histone demethylase LSD1 and to identify regulatory sequences affected by LSD1 depletion.

    \item \textbf{GSE295902 (EBF1):} Publicly available ATAC-seq datasets from mouse (\textit{Mus musculus}) pro-B cells were used to construct training and evaluation datasets for the convolutional neural network (CNN) model. The dataset originates from a study examining the epigenetic regulation of B-cell lineage commitment mediated by EBF1 and PAX5, providing genome-wide chromatin accessibility profiles under WT-type and EBF1-deficient conditions. Differentially accessible regions between EBF1 WT and EBF1-KO cells were identified and used to define sequence classes for model training (\url{https://www.ncbi.nlm.nih.gov/geo/query/acc.cgi?acc=GSE295902}) \cite{36}.
\end{itemize}
To generate sequence datasets representing TF-dependent and TF-independent regulatory elements, ATAC-seq peaks from WT and KO samples were first filtered to obtain unique, non-redundant peak regions within each dataset. Subsequently, a genome-interval comparison was performed to identify differential peak sets. Peaks detected only in WT samples were classified as accessible WT regions, while peaks detected only in KO samples were classified as accessible KO regions. The genomic coordinates of these specific peaks were used to extract fixed-length DNA sequences centered on the peak summits of the mm10 reference genome. These sequences were transformed into numerical representations through one-hot encoding and used as input for CNN-based classification and motif discovery.
The final distribution of genomic sequences utilized for model training and validation of these two datasets is added in Table 1 with GEO ID, number of WT and KO samples, and target factor. All these datasets are added in Supplementary File S1.
\begin{table}[htbp]
\centering
\caption{Summary of Experimental Datasets and Sample Sizes}
\label{tab:datasets}
\begin{tabular}{|c|c|c|c|c|}
\hline
\textbf{GEO ID} & \textbf{Target Factor} & \textbf{WT ($n$)} & \textbf{KO ($n$)} & \textbf{Total ($N$)} \\ \hline
GSE107075 & RelA (p65) & 3722 & 6164 & 9886\\ \hline
GSE119222 & LSD1 & 315 & 733 & 1,048 \\ \hline

GSE295902 & Ebf1 & 42230 & 16725 & 58955 \\ \hline

\end{tabular}
\end{table}


\subsection{CNN-Based Peak Prediction and Discriminative K-mer Extraction}
\begin{figure}
    \centering
    \fbox{\includegraphics[width=1\textwidth]{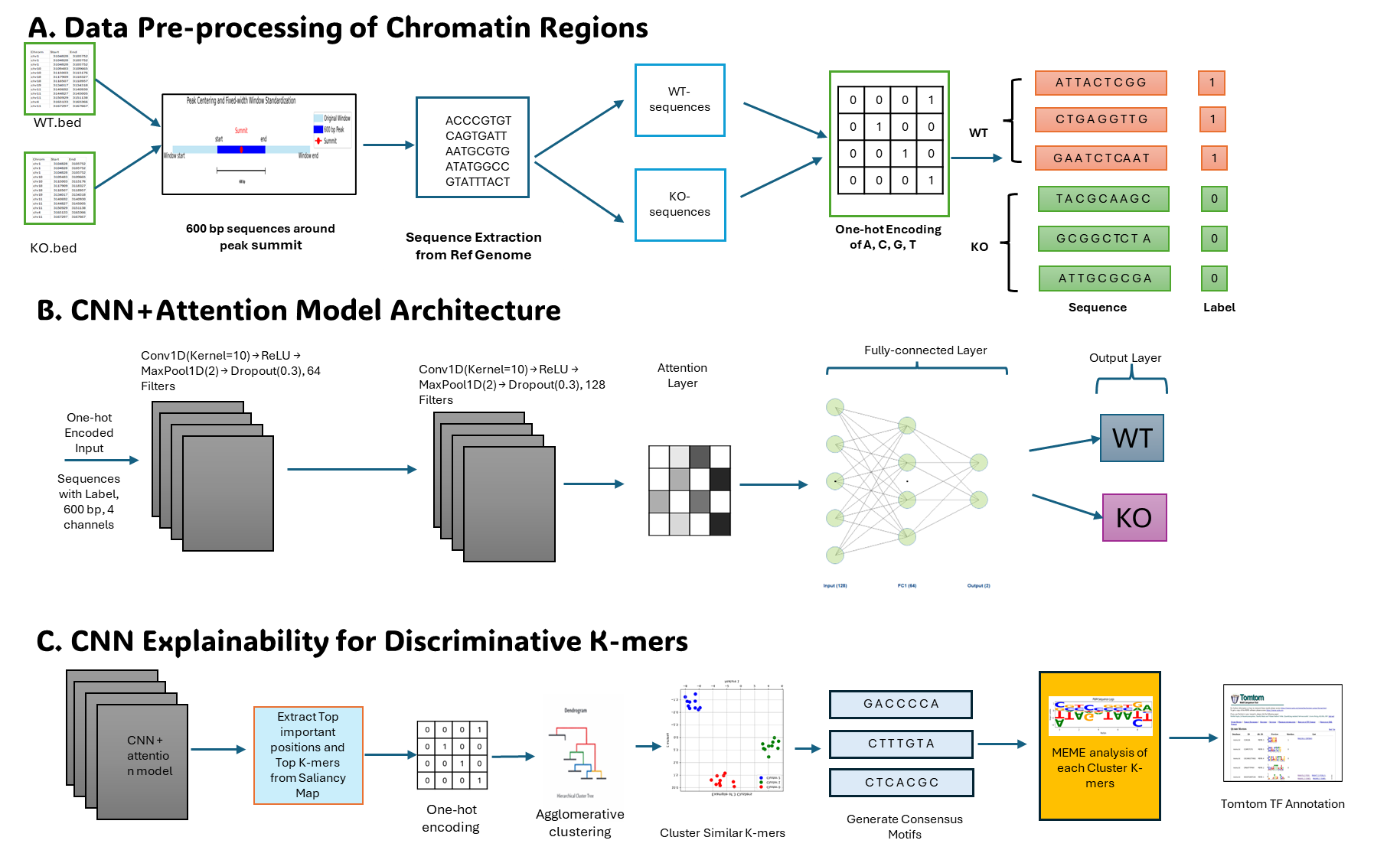}}
    \caption{Schematic overview of the CNN-based pipeline for WT and KO ATAC-seq peak classification.
(A) Data preprocessing steps, including sequence extraction of peaks from reference genome, one-hot encoding of DNA sequences, and class labelling of WT- and KO-specific peaks.
(B) Convolutional neural network architecture for supervised classification of WT- and KO-specific ATAC-seq peaks based on sequence features for peak prediction.
(C) Saliency-guided k-mer extraction and motif analysis. Saliency maps derived from the trained CNN are used to identify the most informative nucleotide positions within each sequence. Twenty-base-pair k-mers centered on high-saliency positions are extracted and subsequently clustered using agglomerative clustering to group similar sequence patterns. Consensus motifs are generated from each cluster using MEME, and the resulting motifs are compared against known transcription factor databases using TOMTOM to identify candidate regulatory factors associated with condition-specific chromatin accessibility.}
    \label{fig: 5}
\end{figure}

To investigate whether DNA sequence features alone can distinguish between WT and KO chromatin accessibility, I developed a convolutional neural network (WTKO-CNN) model for binary peak classification. An overview of the CNN-based WT and KO ATAC-seq peak classification pipeline is provided in Figure~\ref{fig: 5}, illustrating the sequential steps from peak preprocessing and one-hot encoding to dataset balancing, model training, and performance evaluation.
\subsubsection{Data Preprocessing: Extraction and Labeling of Sequences}
To prepare genomic data for convolutional neural network (CNN) analysis, raw ATAC-seq peaks are processed to ensure uniform input dimensions. For each differentially accessible region, the summit of the peak defined as the genomic coordinate with the maximum read density was identified. Then a fixed-width regulatory window of 600~bp was defined by centring on the summit ($\text{summit} \pm 300$~bp). This centering strategy ensures that core regulatory features, including transcription factor binding sites (TFBS), which tend to cluster near peak summits, are consistently positioned within the receptive field of the first convolutional layer. All of these RelA, LSD1, and Ebf1 datasets are available in Supplementary File S1.
\begin{table}[h!]
\centering
\begin{tabular}{|c|c|c|}
\hline
\textbf{Position} & \textbf{Base} & \textbf{Encoding} \\ \hline
1 & A & [1, 0, 0, 0] \\ \hline
2 & C & [0, 1, 0, 0] \\ \hline
3 & G & [0, 0, 1, 0] \\ \hline
4 & T & [0, 0, 0, 1] \\ \hline
5 & N & [0, 0, 0, 0] \\ \hline
\end{tabular}
\caption{One-hot encoding of the nucleotide sequence ``ACCCGTCA''}
\end{table}
The resulting genomic windows are extracted from the mm10 reference genome, yielding sequences of uniform length (600~bp). Each sequence was encoded using a one-hot representation with four channels corresponding to the nucleotides A, C, G, and T. Specifically, adenine (A) was encoded as [1, 0, 0, 0], cytosine (C) as [0, 1, 0, 0], guanine (G) as [0, 0, 1, 0], and thymine (T) as [0, 0, 0, 1]. Ambiguous nucleotides (N) are represented as [0, 0, 0, 0] (Table~2). An example of one-hot encoding for the sequence ``ACCCGTCA'' is shown in Figure~2.
\begin{figure}[h]
\centering
\begin{tikzpicture}

\node at (0, 0) {\textbf{} ACCCGTCA};

\draw[->, thick] (1.5, 0) -- (3, 0);

\node at (5, 0) {
    \begin{math}
    \left[ \begin{array}{cccc}
    1 & 0 & 0 & 0 \\
    0 & 1 & 0 & 0 \\
    0 & 0 & 1 & 0 \\
    0 & 0 & 0 & 1 \\
    1 & 0 & 0 & 0 \\
    0 & 1 & 0 & 0 \\
    0 & 0 & 1 & 0 \\
    0 & 0 & 0 & 1 \\
    \end{array} \right]
    \end{math}
};
\end{tikzpicture}
\caption{Illustration of sequence "ACCCGTCA" being transformed into a one-hot encoded matrix and fed into a CNN layer for further processing.}
\end{figure}
The one-hot encoded sequence is treated by the CNN as a two-dimensional input matrix (e.g., $8 \times 4$ for the illustrative example), where rows correspond to nucleotide positions and columns correspond to nucleotide identities. Convolutional filters (kernels) scan this matrix to learn local sequence patterns and regulatory motifs relevant to chromatin accessibility. 

As the downstream task involved binary peak classification, each sequence was assigned a class label based on condition-specific chromatin accessibility. Peaks specific to the wild-type (WT) condition are assigned label 1, whereas peaks specific to the knockout (KO) condition are assigned label 0. The one-hot encoded sequences and their corresponding labels are used as inputs for training the convolutional neural network (CNN) model. To avoid bias during training, class imbalance between WT- and KO-specific peaks was corrected by balancing the dataset according to peak counts, resulting in an equal number of sequences for each class used in supervised learning.

The CNN architecture was designed with two convolutional layers to hierarchically extract regulatory sequence features from the one-hot encoded DNA inputs. Because peak lengths varied substantially across the dataset (ranging from 150 to 2567 bp), all sequences were standardized to a fixed length of 600 bp before training. This length was empirically determined using saliency map analysis, which indicated that the most informative and discriminative sequence features are typically concentrated within an approximately 600 bp window centered on the peak summit. Standardizing sequence length ensured consistent input dimensions for the CNN while preserving the most biologically relevant signal.

The convolutionally generated feature maps are subjected to an attention layer to further identify physiologically significant regions within each sequence. In order to create attention weights that indicate the relative significance of each position in the sequence, an attention score was calculated for each position and normalised using the softmax algorithm. The model was able to concentrate on the most informative areas for classification since the final sequence representation was produced as a weighted sum of positional feature vectors.

\subsubsection{Training and Test Split}
Following sequence preprocessing, one-hot encoding, and dataset balancing, I have implemented a supervised convolutional neural network (CNN) to classify ATAC-seq peaks as WT- or KO-specific based solely on their underlying DNA sequence. To prevent data leakage, sequences from chromosomes 6 and 7 are held out as a test set, while sequences from the remaining chromosomes are used for training. This ensured that the model was evaluated on completely independent, unseen genomic regions, which is essential for a fair and realistic assessment of model performance and further use of the model for peak prediction for other unseen regions. 
\subsubsection{CNN Architecture}
Each input sequence was represented as a $600 \times 4$ matrix, where rows corresponded to nucleotide positions and columns to nucleotides (A, C, G, T). The CNN architecture consisted of two one-dimensional convolutional layers. The first layer contained 64 filters with a kernel size of 10 bp, followed by batch normalization, ReLU activation, max-pooling, and dropout (rate = 0.3). The second convolutional layer contained 128 filters (kernel size = 10 bp), with similar batch normalization, ReLU, max-pooling, and dropout. This design allows the network to learn motif-like sequence features while reducing overfitting.

The output of the convolutional layers was flattened and passed through a fully connected layer of 64 neurons, followed by a two-node output layer with a softmax activation to predict WT or KO class probabilities. The model was trained using the Adam optimizer with an initial learning rate of 0.001, L2 weight decay (0.0001), and binary cross-entropy loss. Mixed precision training and gradient accumulation (steps = 4) are used to improve efficiency and stability. Training was performed for 100 epochs, with the best model selected based on maximum test set accuracy (Table 3).
\subsubsection{Model Training}
The model was implemented in PyTorch and trained using the Adam optimizer with a learning rate of $3\times10^{-4}$ and weight decay of $1\times10^{-4}$. Cross-entropy loss was used as the optimization objective. Training was performed for 100 epochs with gradient accumulation to simulate larger batch sizes. Mixed-precision training was enabled when GPU hardware was available. 

\subsubsection{Model Evaluation}
After training, the best-performing model was used to predict labels for the test set. Performance was assessed using overall accuracy, confusion matrix, and classification metrics including precision, recall, and F1-score for each class. Model performance was monitored per epoch, tracking both training and test accuracy. This pipeline enabled identification of sequence features predictive of chromatin accessibility differences between WT and KO conditions. The best model checkpoint was saved for downstream evaluation. The CNN architecture used in this study is summarized in the following pseudocode. The model integrates convolutional feature extraction with an attention mechanism for sequence-level classification.

\begin{lstlisting}[caption=Pseudocode for CNN-based Chromatin Accessibility Classification]

Input: WT and KO DNA sequences (variable length, padded to 600 bp)
Labels: WT = 1, KO = 0

\begin{lstlisting}[caption=CNN + Attention Model for Chromatin Accessibility Classification]

Input: WT/KO DNA sequences (padded to 600 bp), labels (WT=1, KO=0)

1. Data:
- Balance WT and KO samples
- Split by chromosome:
    Train: chromosomes 1-19 Except Chromosome 6 and 7
    Test: chromosome 6 and 7
- One-hot encode (A,C,G,T,N) → (600 × 4)

2. Model:
- Input → reshape to (batch, 4, 600)
- Conv1D(64, kernel=10) + ReLU + MaxPool + Dropout(0.3)
- Conv1D(128, kernel=10) + ReLU + MaxPool + Dropout(0.3)
- Attention over sequence dimension
- Dense(128->64) + ReLU
- Dense(64->2 logits)

3. Training:
- Loss: CrossEntropyLoss
- Optimizer: Adam (lr=0.001)
- Mixed precision (autocast + GradScaler)
- Gradient accumulation (4 steps)
- Train for 100 epochs

4. Evaluation:
- Test on chr6/chr7 each epoch
- Save best model by test accuracy
- Final metrics: confusion matrix, precision, recall, F1

5. Output:
- Accuracy curves
- Confusion matrix
- Best model weights

6. Interpretation:
   - Saliency maps
\end{lstlisting}

\begin{table}[h]
\centering
\caption{Architecture of the 1D Convolutional Neural Network with Attention for sequence classification.}
\label{tab:model_arch_attention}
\begin{tabular}{@{}llll@{}}
\textbf{Layer} & \textbf{Type} & \textbf{Parameters} & \textbf{Activation} \\ \hline
Input & Sequence & $600 \times 4$ nucleotides & --- \\
Conv1D (1) & Convolutional & 64 filters, kernel size=10 & ReLU \\
Max Pool (1) & Max Pooling & pool size=2 & --- \\
Dropout (1) & Regularization & $p=0.3$ & --- \\
Conv1D (2) & Convolutional & 128 filters, kernel size=10 & ReLU \\
Max Pool (2) & Max Pooling & pool size=2 & --- \\
Dropout (2) & Regularization & $p=0.3$ & --- \\
Attention & Attention & 128-dimensional feature vectors & Softmax weights \\
Fully Connected & Linear & 64 units & ReLU \\
Output & Linear & 2 units (Logits) & --- \\ \hline
\end{tabular}
\end{table}
\subsection{Saliency-Based Identification of Regulatory Sequence Features}

To interpret the regulatory sequence patterns learned by the WTKO-CNN model, I applied gradient-based attribution using saliency mapping to estimate the contribution of individual nucleotide positions to the model's predictions. For a given input sequence $X \in \{0,1\}^{L \times 4}$, where $L = 600$, the saliency map $S(X)$ was defined as the gradient of the predicted class logit $f_c(X)$ (pre-softmax) with respect to the input sequence:

\begin{equation}
S(X) = \frac{\partial f_c(X)}{\partial X}
\end{equation}

The magnitude of this gradient reflects the local sensitivity of the model’s prediction to perturbations at each nucleotide position. Because DNA sequences are represented using one-hot encoding, gradients are computed across the four nucleotide channels $j \in \{A, C, G, T\}$. To obtain a single positional importance score $M_i$, we selected the maximum absolute gradient value across the nucleotide channels:

\begin{equation}
M_i = \max_{j \in \{A, C, G, T\}} |S_{i,j}|
\end{equation}

This operation produced a position-wise importance profile for each sequence, where nucleotides with the highest saliency scores are interpreted as the most influential features contributing to the classification decision.

For each sequence in the test dataset (chromosomes 6 and 7), the three positions with the highest $M_i$ values are identified as saliency peaks. Positions with the highest saliency scores are interpreted as the most influential regions contributing to the classification decision. For each sequence in the test dataset, the top three highest-scoring positions are identified. To capture potential regulatory motifs, 20-bp  $k$-mers centered on each peak position $p_n$, centered around these high-importance positions are extracted. 

\begin{equation}
k\text{-mer}_n = X[p_n - 6 : p_n + 7]
\end{equation}

The potential regulatory motifs that underlie the differences in accessibility between WT and KO situations are represented by these extracted subsequences. For further analysis and visualisation, these motif candidates are gathered. Saliency maps are also created for each sequence to show how positional importance is distributed across the sequence and to emphasise the areas that have the biggest impact on the model's predictions.

\subsection{Motif Discovery from Saliency-Identified k-mers}
To identify the sequence motifs learned by the WTKO-CNN model, top k-mers (20bp) centered on saliency peaks are first extracted from each sequence. Each sequence contributed multiple k-mers corresponding to its highest-saliency positions. These k-mers are then flattened into a single table, with each row representing one k-mer and its associated sequence index and position.

The flattened k-mers are numerically encoded using one-hot representation for the four standard nucleotides (A, C, G, T), with ambiguous positions treated as a separate channel.  Next, these k-mers are clustered using agglomerative hierarchical clustering. We used cosine distance and average linkage, to group similar sequence patterns into clusters. For each cluster, a consensus motif and a weighted position matrix are computed. Then MEME is applied on each cluster k-mer fasta sequences separately. 

The motifs identified by MEME are compared with known transcription factor binding site (TFBS) databases using TOMTOM, allowing us to associate CNN-derived sequence features with established regulatory elements. This workflow enabled systematic interpretation of the sequence features driving model predictions and facilitated direct comparison with known TFBS. Figure 3 illustrates the workflow for saliency-based motif discovery from the WTKO-CNN model.
\begin{figure}
    \centering
    \includegraphics[width=1\linewidth]{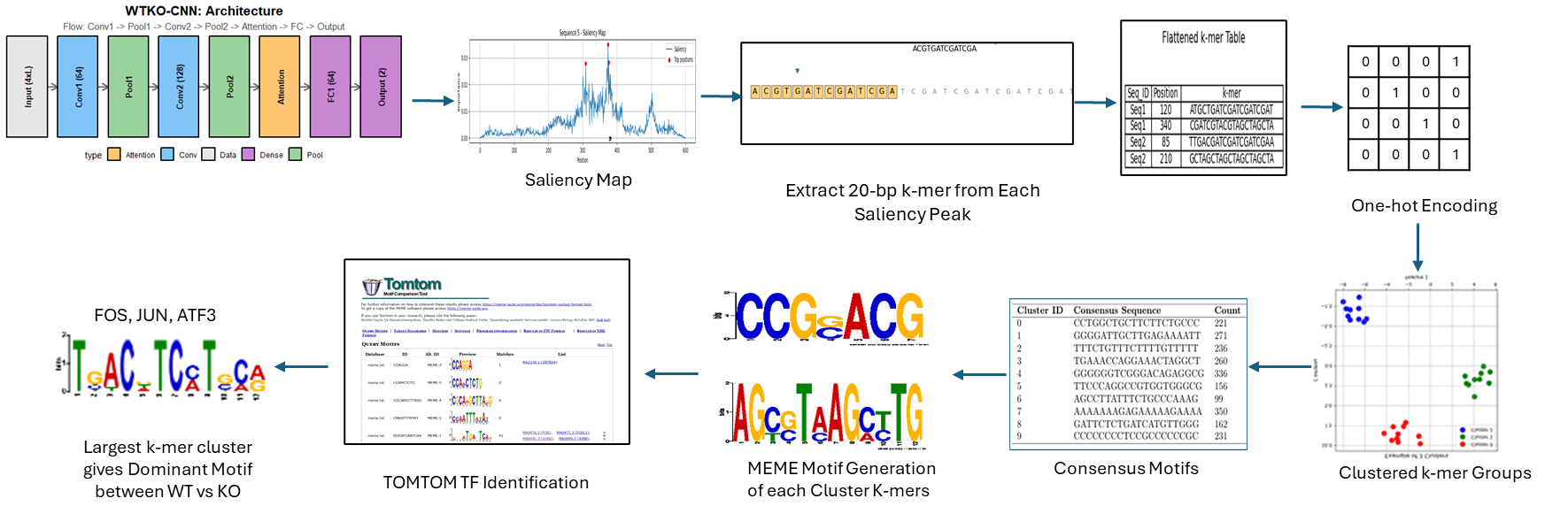}
    \caption{The workflow for motif discovery from the WTKO-CNN model, including saliency-based k-mer extraction, clustering, motif generation, and comparison with known TFBS using MEME and TOMTOM.}
    \label{fig:placeholder}
\end{figure}
\subsection{Identification of Discriminative $k$-mers}
Importantly, in this work, I applied explainable AI (XAI) methods to learn more about how our CNN model makes decisions. To ensure the model learned generalizable biological motifs rather than locus-specific features, I implemented a strict genomic partition. Filters are optimized using sequences from the majority of the genome, while Chromosomes 6 and 7 are held out as an independent test set. This spatial separation ensures that the motifs identified by our convolutional filters represent conserved regulatory signatures across the entire murine genome. I extracted attention-weighted k-mers from the model to identify candidate regulatory elements enriched in either WT or KO peaks. This method not only helped us identify the sequence sections that are essential to the model's predictions, but it also made it possible to link these significant regions to motifs that had biological significance. I found k-mers that are aggregated from saliency scores using a sliding window approach, and ranked them according to how relevant they are to the classification job. 

To identify sequence motifs associated with WT- and KO-specific chromatin accessibility, I extracted k-mers from the first convolutional layer of our trained CNN models. I analyzed the weights of the first convolutional layer. Each of the 64 filters represents a position weight matrix (PWM) of length $k=10$. We converted these weights into consensus $k$-mers by identifying the nucleotide with the maximum weight at each position:
\begin{equation}
    S_j = \text{argmax}_{i \in \{A,C,G,T\}} W_{i,j}
\end{equation}
where $W_{i,j}$ represents the weight of nucleotide $i$ at position $j$. Given that random weight initialization and dropout can result in slight variability in learned filters, I trained the network independently five times, each with a unique random seed. For each run, high-activation subsequences corresponding to the convolutional filters are collected as candidate k-mers.  

To ensure robust motif identification, I aggregated k-mers across all five runs and retained only those that are consistently detected in multiple models, thereby emphasizing recurrent and biologically relevant patterns. Extracted k-mers are subsequently analyzed using motif discovery tools, allowing comparison with known transcription factor binding sites and identification of novel regulatory elements underlying differential chromatin accessibility between WT and KO conditions. 

The model was shown to have learned biologically important patterns when the top-ranked k-mers are compared with known transcription factor binding motifs from JASPAR and HOMER. The interpretability of our CNN model is further demonstrated by the identification of unique motifs that are absent from both databases, indicating the approach's potential to find new regulatory components. In addition to validating the model's predictions, this combination of XAI approaches bridges the gap between machine learning and biological knowledge by offering a deeper understanding of the chromatin accessibility landscape. Figure 4 shows the Interpretability Pipeline for Genomic Sequence Classification. 

\begin{figure}
    \centering
    \includegraphics[width=0.8\linewidth]{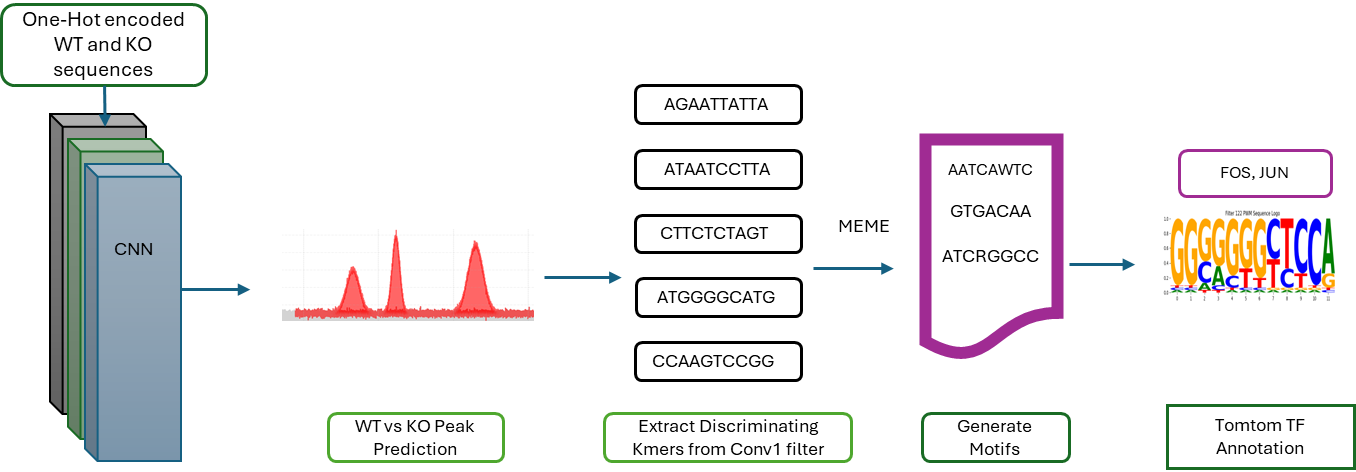}
    \caption{\textbf{Schematic of the Interpretability Pipeline for Genomic Sequence Classification}. One-hot encoded Wild Type (WT) and Knockout (KO) sequences are fed into a deep learning model to classify the peak. To interpret the model, discriminating k-mers are extracted from the learned filters. These sequences are processed using MEME for motif discovery, and the resulting motifs are annotated with transcription factor (TF) identities using Tomtom.}
    \label{fig:placeholder}
\end{figure}
.




\section{Results}
\subsection{Model Training and Performance}
To evaluate the ability of WTKO-CNN to capture sequence determinants of chromatin accessibility, we trained the model on genome-wide ATAC-seq data while holding out chromosomes 6 and 7 for independent testing. This chromosome-level holdout ensures that performance reflects the model’s ability to generalize to unseen genomic regions, rather than memorizing training sequences. On the held-out chromosomes, WTKO-CNN accurately distinguished WT from KO peaks, achieving, demonstrating robust predictive performance across the genome. The CNN was trained to classify WT- and KO-specific ATAC-seq peaks based on 600~bp one-hot encoded sequences. Training was performed for 100 epochs with gradient accumulation and mixed precision. Early epochs showed a steady increase in both training and test accuracy. Beyond this point, the training accuracy continued to increases, while test accuracy decreases, indicating overfitting .  

\begin{figure}[htbp]
    \centering
    \begin{subfigure}[b]{0.45\textwidth}
        \centering
        \includegraphics[width=\linewidth]{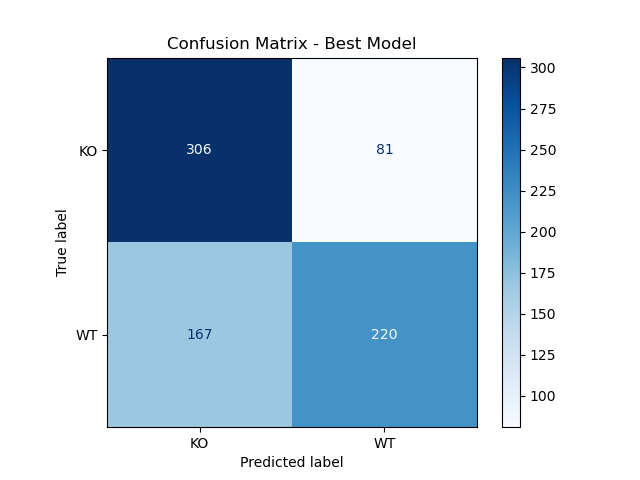}
        \caption{GSE107075 (RelA)}
        \label{fig:sub1}
    \end{subfigure}
      \begin{subfigure}[b]{0.45\textwidth}
        \centering
        \includegraphics[width=\linewidth]{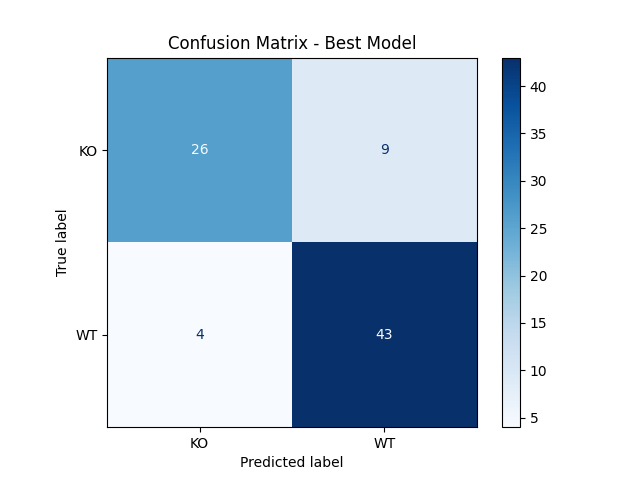}
        \caption{GSE119222 (LSD1)}
        \label{fig:sub2}
    \end{subfigure}
     \begin{subfigure}[b]{0.45\textwidth}
        \centering
        \includegraphics[width=\linewidth]{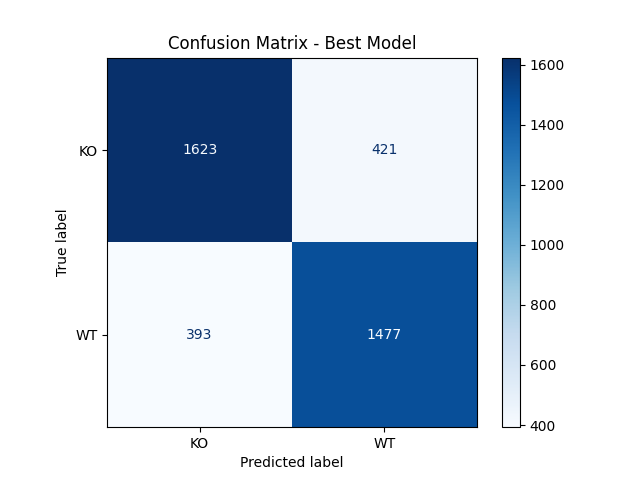}
        \caption{GSE295902 (Ebf1)}
        \label{fig:sub2}
    \end{subfigure}
            
    \caption{WT vs KO Peak Prediction. Training is done using chromosomes 1 to 19, except chromosomes 6 and 7 and Testing using chromosomes 6 and 7.}
    \label{fig:all_images}
\end{figure}

Due to stochastic weight initialization and data shuffling, model performance can vary slightly across runs. To ensure robustness, I trained the model multiple times and reported the best-performing model results.  WTKO-CNN was evaluated on three independent ATAC-seq datasets (GSE107075, GSE119222, GSE295902) to assess its generalization beyond the training data. As shown in Table~\ref{tab:wtko-cnn-performance}, the model achieved moderate to high accuracy, ranging from 68\% to 84\%, with per class precision, recall, and F1-scores reflecting variability between WT and KO peaks. Multiple random seeds were evaluated for each dataset, and the results reported correspond to the seed achieving the highest validation performance. Figure~\ref{fig:all_images} shows the confusion matrix of the WTKO-CNN prediction. Differences in performance across datasets likely arise from a combination of factors, including technical variations in ATAC-seq protocols, biological heterogeneity of the samples, and subtle differences in the magnitude of chromatin accessibility changes between WT and KO conditions. Despite these variations, WTKO-CNN consistently identified informative k-mer patterns corresponding to transcription factor binding motifs enriched in condition-specific peaks, demonstrating its ability to capture biologically meaningful regulatory features even in external datasets. All experiments were executed on an NVIDIA GeForce RTX 4070 GPU with CUDA acceleration.
\begin{table}[htbp]
\centering
\caption{Performance of WTKO-CNN on held-out chromosomes and training data.}
\label{tab:wtko-cnn-performance}
\begin{tabular}{|l|l|l|l|l|l|l|}
\hline
\textbf{Dataset} & \textbf{Seed} & \textbf{Class} & \textbf{Precision} & \textbf{Recall} & \textbf{F1-score} & \textbf{Accuracy} \\ \hline
  RelA (GSE107075)   & 30     & WT & 0.73 & 0.57 & 0.64 & 68\% \\ 
         & & KO & 0.65 & 0.79 & 0.71 &  \\ \hline
LSD1 (GSE119222)  & 20    & WT & 0.83  & 0.91 & 0.87 & 84\% \\ 
 & & KO & 0.87 & 0.74 & 0.80 &  \\ \hline
 Ebf1 (GSE295902)  & 30  & WT & 0.78  & 0.79 & 0.78 & 79\% \\ 
  &  & KO & 0.81 & 0.79 & 0.80 &  \\ \hline

\end{tabular}
\end{table}

\begin{figure}[htbp]
\centering
\begin{subfigure}{0.49\textwidth}
\includegraphics[width=\linewidth]{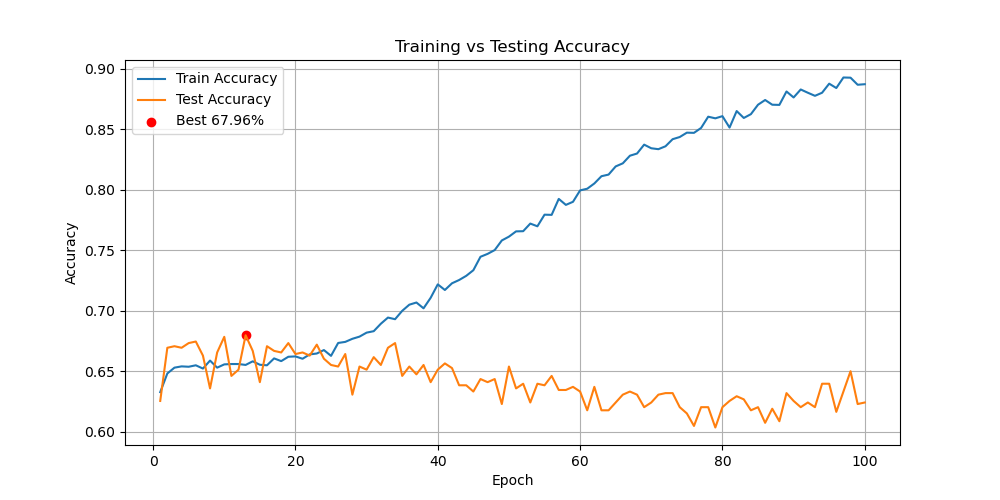}
\caption{GSE107075 (RelA)}
\label{fig:accuracy_d1}
\end{subfigure}
\hfill
\begin{subfigure}{0.49\textwidth}
\includegraphics[width=\linewidth]{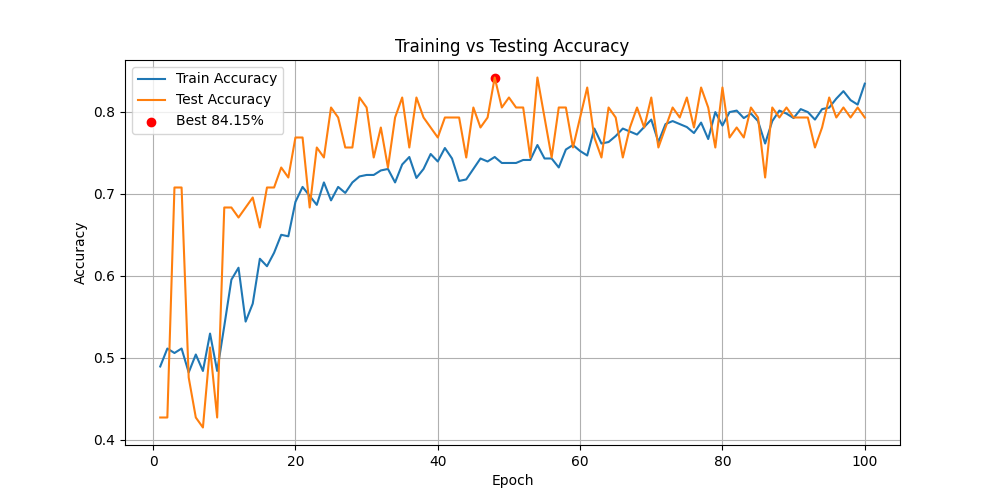}
\caption{GSE119222 (LSD1)}
\label{fig:accuracy_d2}
\end{subfigure}
\begin{subfigure}{0.49\textwidth}
\includegraphics[width=\linewidth]{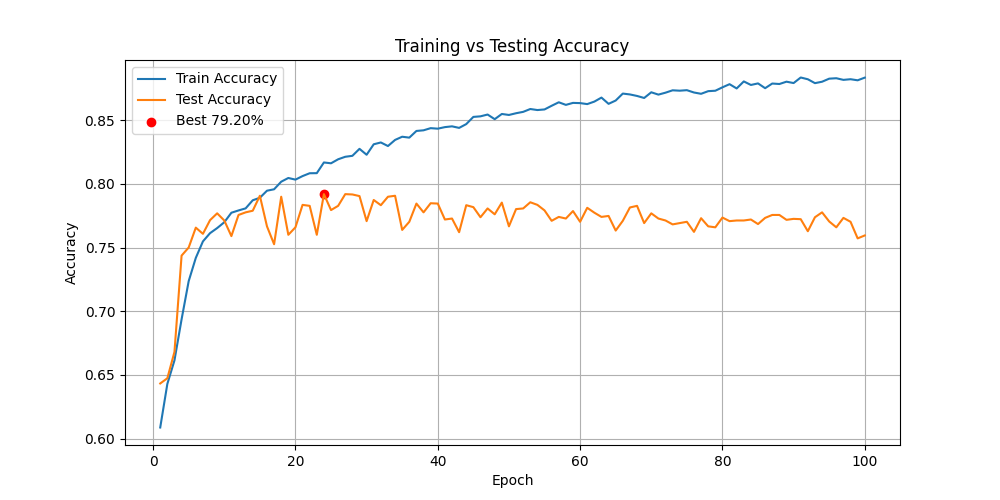}
\caption{GSE295902 (Ebf1)}
\label{fig:accuracy_d2}
\end{subfigure}
\caption{Training and test accuracy curves per epoch for RelA, LSD1 and Ebf1 datasets. 
For each dataset, the CNN--attention model’s training (blue) and test (orange) accuracy are shown. Red dots indicate epochs with the highest test accuracy.}
\label{fig:accuracy_all}
\end{figure}

Figure~\ref{fig:accuracy_all} shows the training (blue) and test (orange) accuracy curves across all epochs. Both curves gradually increased and stabilized, with the test accuracy. The model achieving the highest test accuracy was saved as the final model for downstream saliency and motif analyses. The small gap between training and test accuracy indicates that the model generalized well without substantial overfitting.

\subsection{Saliency-Based Identification of Regulatory Sequence Features}

\begin{figure}[htbp]
\centering

\begin{subfigure}{0.49\textwidth}
\centering
\includegraphics[width=\linewidth]{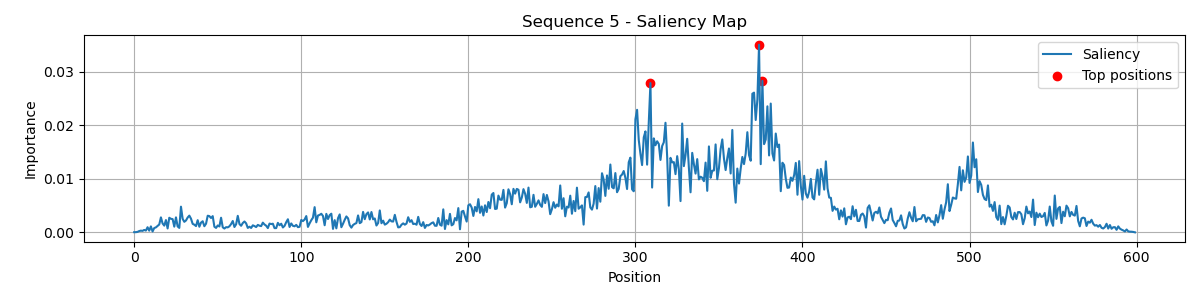}
\caption{RelA-Sequence 5}
\end{subfigure}
\hfill
\begin{subfigure}{0.49\textwidth}
\centering
\includegraphics[width=\linewidth]{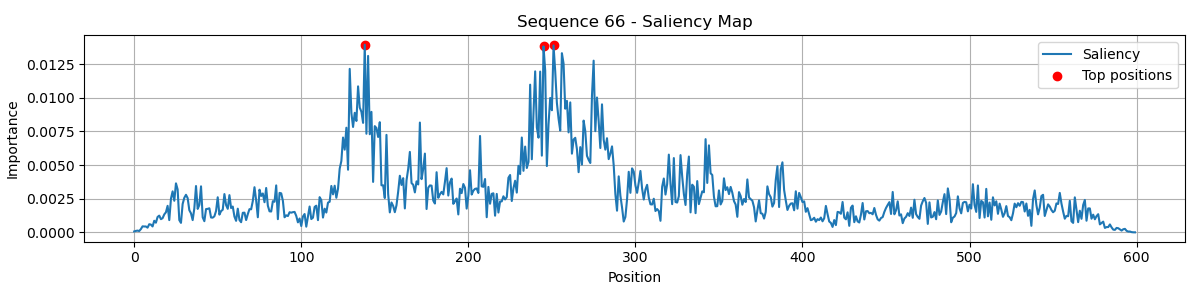}
\caption{RelA-Sequence 66}
\end{subfigure}

\vspace{0.3cm}

\begin{subfigure}{0.49\textwidth}
\centering
\includegraphics[width=\linewidth]{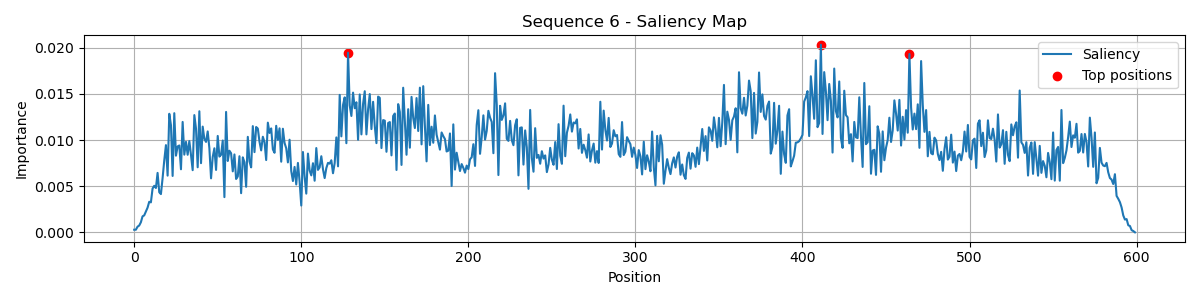}
\caption{LSD1-Sequence 6}
\end{subfigure}
\hfill
\begin{subfigure}{0.45\textwidth}
\centering
\includegraphics[width=\linewidth]{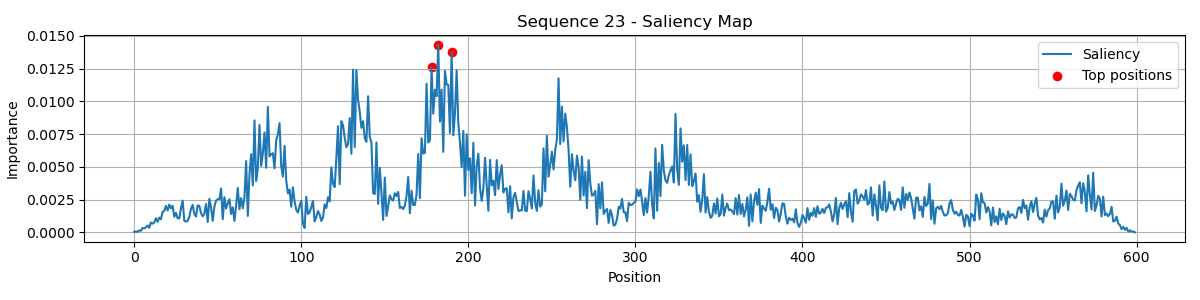}
\caption{LSD1-Sequence 23}
\end{subfigure}
\vspace{0.3cm}

\begin{subfigure}{0.49\textwidth}
\centering
\includegraphics[width=\linewidth]{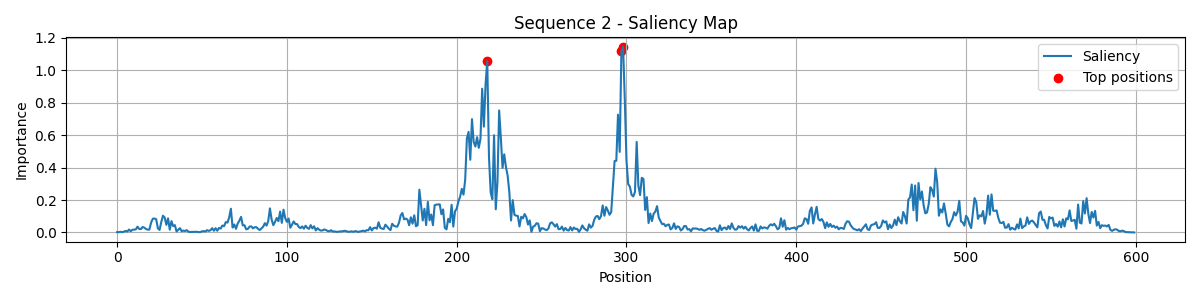}
\caption{Ebf1-Sequence 2}
\end{subfigure}
\hfill
\begin{subfigure}{0.45\textwidth}
\centering
\includegraphics[width=\linewidth]{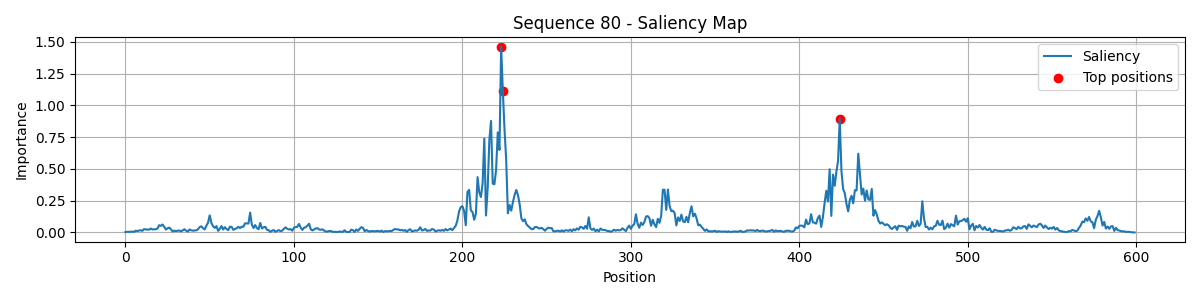}
\caption{Ebf1-Sequence 80}
\end{subfigure}
\caption{Sample saliency maps illustrating nucleotide-level importance across test genomic sequences of RelA, LSD1, and Ebf1 Dataset. Peaks in saliency highlight regions contributing most strongly to model predictions.}
\label{fig:saliency_examples}

\end{figure}

To interpret the features driving CNN attention predictions, I computed gradient-based saliency maps for each sequence in the test set (Chr6 and Chr7). The three positions with the highest saliency scores are selected per sequence, and 20-bp k-mers centered on these positions are extracted as candidate regulatory elements (Figure~\ref{fig:saliency_examples}). Saliency analysis revealed that model predictions are influenced by specific sequence motifs rather than distributed noise across the input, indicating that the model captured biologically meaningful patterns.

Sample saliency maps illustrating nucleotide-level importance for model predictions are shown in Figure~\ref{fig:saliency_examples}. For each dataset, two sequences from the held-out test set are selected to visualize the positional contribution of nucleotides to the classification decision. Regions with high saliency values indicate sequence positions that most strongly influenced the CNN--attention model.

\subsection{Motif Discovery and Enrichment Analysis}
In this step, I flattened the top k-mers extracted from high-saliency positions into a single table and clustered using hierarchical agglomerative clustering based on cosine distance to group similar sequence patterns. For each cluster, we generated a consensus motif by selecting the most frequent nucleotide at each position, and then de novo motif discovery was performed using MEME \cite{31}. Further, MEME-generated motifs are compared against known transcription factor binding sites using TOMTOM. Clusters with fewer than 10 k-mers are not analyzed with MEME/TOMTOM due to insufficient sequence numbers for reliable motif inference. A list of consensus motifs are given in Table~\ref{tab:consensus_motifs_ds1} for RelA, Table~\ref{tab:lsd1_motifs} for LSD1 and Table~\ref {tab:consensus_motifs_ebf1} for Ebf1. I considered 10 clusters for RelA and Ebf1. As the number of test sequences for LSD1 was smaller, we considered 6 clusters for LSD1. 

Top transcription factor matches and their respective families are annotated based on TOMTOM results are shown in Table~\ref{tab:MEME_TF_matches} and Table~\ref{tab:MEME_TF1_matches}. This analysis revealed that many high-saliency clusters correspond to biologically relevant motifs, including members of the bZIP family (FOS, JUNB, BATF), bHLH (ASCL1, TCF4) and C2H2 zinc finger factors (KLF4, KLF5), providing mechanistic insight into the regulatory grammar learned by the model. Table~\ref{tab:MEME_TF_matches} and Table~\ref{tab:MEME_TF1_matches} summarizes representative clusters, including consensus motifs, motif logos, top transcription factor matches, and their families of RelA dataset. This cluster-wise approach allowed identification of both common and condition-specific motifs, reflecting differential regulatory patterns between WT and KO sequences.

Clusters containing at least 10 k-mers are considered for motif analysis. Among these, the cluster with the largest number of k-mers are treated as the primary discriminative transcription factor (TF) between WT and KO, while the remaining clusters are considered secondary discriminative TFs. Cluster 4 for RelA, Cluster 1 for LSD1 and Cluster 5 for Ebf1 contained the most k-mers and are treated as primary discriminative clusters, with the remaining clusters considered secondary. It can be seen from the tables, bZIP family (FOS, JUN, ATF3) acts as the primary discriminative TF for RelA, whereas bHLH (ASCL1, TCH4) family for LSD1. Interestingly, For the Ebf1 dataset, CNN-derived motif clustering identified both EBF1- and RUNX-associated sequence patterns as discriminative features between WT and KO conditions. The EBF1-associated motif cluster (cluster 0; consensus sequence CCACACTGACCCCACAGGGC) showed 2,194 occurrences, whereas the RUNX-associated motif cluster (cluster 5; consensus sequence GGTCCCCGGGGGTTTTGGGG) displayed a higher frequency with 6,312 occurrences among the extracted high-activation k-mers. The increased representation of RUNX-associated motifs in the discriminative sequence features suggests a potential shift toward alternative regulatory elements following Ebf1 perturbation. These findings raise the possibility that RUNX family transcription factors may contribute to compensatory regulation or maintain regulatory activity in Ebf1 KO conditions. All TOMTOM results are added in Supplementary Folder S4.

\begin{table}[h]
\centering
\caption{Consensus motifs derived from clustered high-saliency k-mers for RelA Dataset. Each cluster groups k-mers with similar sequences, and the count indicates the number of k-mers in that cluster.}
\label{tab:consensus_motifs_ds1}
\begin{tabular}{lll}
\hline
\textbf{Cluster ID} & \textbf{Consensus Sequence} & \textbf{Count} \\
\hline
0 & CCTGGCTGCTTCTTCTGCCC & 221 \\
1 & GGGGATTGCTTGAGAAAATT & 271 \\
2 & TTTCTGTTTCTTTTGTTTTT & 236 \\
3 & TGAAACCAGGAAACTAGGCT & 260 \\
4 & GGGGGGTCGGGACAGAGGCG & 336 \\
5 & TTCCCAGGCCGTGGTGGGCG & 156 \\
6 & AGCCTTATTTCTGCCCAAAG & 99 \\
7 & AAAAAAAGAGAAAAAGAAAA & 350 \\
8 & GATTCTCTGATCATGTTGGG & 162 \\
9 & CCCCCCCCTCCGCCCCCCGC & 231 \\
\hline
\end{tabular}
\end{table}
\begin{table}[ht]
\centering
\caption{Consensus motifs obtained from clustering saliency-derived k-mers for LSD1 binding sequences. Each cluster groups k-mers with similar sequences, and the count indicates the number of k-mers in that cluster.}
\begin{tabular}{lll}
\hline
\textbf{Cluster ID} & \textbf{Consensus Sequence} & \textbf{Count} \\
\hline
 0 & GCCCGGACCGGCTGCCGGGC & 40 \\
 1 & CCGCCCCCCCCCCCCCCCCC & 111 \\
 2 & GGCGCCCCGCCCGCACAGCG & 15 \\
 3 & AAAGTCTGTCGTAGATTCAG & 3 \\
 4 & TCCGGAGCAACAGGGAGAAG & 11 \\
 5 & GGGAGGCAGGCGGGCCCGCA & 63 \\
\hline
\end{tabular}

\label{tab:lsd1_motifs}
\end{table}
\begin{table}[h]
\centering
\caption{Consensus motifs derived from clustered high-saliency k-mers for Ebf1 Dataset. Each cluster groups k-mers with similar sequences, and the count indicates the number of k-mers in that cluster.}
\label{tab:consensus_motifs_ebf1}
\begin{tabular}{lll}
\hline
\textbf{Cluster ID} & \textbf{Consensus Sequence} & \textbf{Count} \\
\hline
0 & CCACACTGACCCCACAGGGC & 2194\\
1 & CCCCCAGTCCCCAGGGGGCC & 1802 \\
2 & TACCAACCACAGGCGGCCCC & 283 \\
3 & GGGTGTTGCCCCACCTTGTG & 88  \\
4 & ACACTAACCACAGACACCAC & 371\\
5 & GGTCCCCGGGGGTTTTGGGG & 6312 \\
6 & ACAGGAAGGCCGCCTCTCTA & 31 \\
7& CTGTGTGGTGAGAGAAAGCC& 29 \\
8& CCACCTGTGGTTTCTGGTGA& 622 \\
9& CATAGCTAACGATTTTTGAG& 10 \\

\hline
\end{tabular}
\end{table}
 
\begin{table}[h!]
\centering
\caption{Summary of cluster-wise MEME motifs derived from high-saliency k-mers and Matched TFs from TOMTOM of RelA Dataset.}
\label{tab:MEME_TF_matches}
\begin{tabular}{|c|c|c|c|c|}
\hline
\textbf{Cluster ID}  & \textbf{MEME Motif} & \textbf{Motif Logo} & \textbf{Top TF match(s)}  & Family\\
\hline
0 & SYCABTGCCTSC & 
\includegraphics[height=0.8cm]{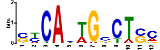}  & 
EHF, ELF3, ELF1 & ETS \\ 
& & & MAFK, NFE2 &  \\ \hline 
1 &	CCACGCGCATCC  & \includegraphics[height=0.8cm]{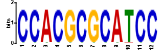} & KLF6, SP8, KLF11, SP9, SP3 & C2H2 zinc finger\\ \hline
1 &  KCCAYGWCTCAG & \includegraphics[height=0.8cm]{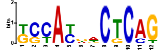} & ATF1,FOS,FOSL2,BACH2 & bZIP \\ \hline
2 & GSBTGWSWGMSH & \includegraphics[height=0.8cm]{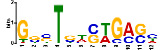} & MAF, MAFA, MAFK & bZIP\\ \hline
4 & MTGASWCA & \includegraphics[height=0.8cm]{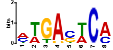} & MAFG, JDP2, BNC2, MAFK & bZIP \\ 

& & & JUNB, JUND, NFE2& \\ \hline
4 & TGACDTCMTSCR & \includegraphics[height=0.8cm]{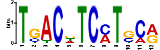} & ATF3, FOS::JUN, FOSL2::JUNB, & bZIP\\ 
& & & ETV1, ELK4, ETS1, FOS, JUN & \\ \hline
5 & TCAGGTTATA & \includegraphics[height=0.8cm]{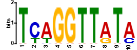} & Six4, SIX1, SIX2 & bZIP\\ \hline
6 & STTTTRGTTK & \includegraphics[height=0.8cm]{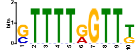} & IRF8, IRF5, IRF4, RUNX1 & Tryptophan cluster,\\
& & & & Runt \\ \hline
7 & GRRACAGWCAT & \includegraphics[height=0.8cm]{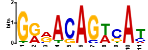} & NFE2, Stat2, FOS::JUND & bZIP \\ \hline
8 & ASTRAGTCAS & \includegraphics[height=0.8cm]{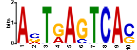} & BNC2, Atf3,BATF::JUN & bZIP \\ \hline
9 & RACTCATCTGA	 & \includegraphics[height=0.8cm]{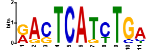} & NFE2, Jun, FOS & bHLH, bZIP\\ \hline
\end{tabular}
\end{table}
\begin{table}[h!]
\centering
\caption{Summary of cluster-wise MEME motifs derived from high-saliency k-mers and Matched TFs from TOMTOM of LSD1 Dataset.}
\label{tab:MEME_TF1_matches}
\begin{tabular}{|c|c|c|c|c|}
\hline
\textbf{Cluster ID}  & \textbf{MEME Motif} & \textbf{Motif Logo} & \textbf{Top TF match(s)}  & Family\\
\hline
0& CCCTGCT & \includegraphics[height=0.8cm]{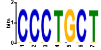} & ZIC3, ZIC1, ZIC2 & C2H2 zinc finger \\
\hline
1 & 	WGNSADCCTGCY & \includegraphics[height=0.8cm]{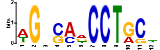} &  TCF4, Zic3, ASCL1, TCF12 & bHLH\\
\hline
1 & AGRBCKGSSKGC & \includegraphics[height=0.8cm]{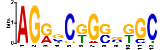}& KLF7, SP2, KLF1 & C2H2 zinc finger\\ 
\hline
2 & CACGTG & \includegraphics[height=0.8cm]{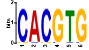} & MAX, Mlxip, MNT & 
bHLH-ZIP \\ 
\hline
4 & AKRAAG & \includegraphics[height=0.8cm]{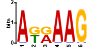} &  Elf5, ELF1, ETV1, EHF, ETS1 & ETS \\ \hline
4 & CMKRRGCAGCWG & \includegraphics[height=0.8cm]{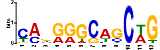} & Ptf1A, Neurod2, Twist2, Fli1  & Tal-related bHLH\\ \hline
\end{tabular}
\end{table}
\begin{table}[h!]
\centering
\caption{Summary of cluster-wise MEME motifs derived from high-saliency k-mers and Matched TFs from TOMTOM of Ebf1 Dataset.}
\label{tab:MEME_TF_matchesebf1}
\begin{tabular}{|c|c|c|c|c|}
\hline
\textbf{Cluster ID}  & \textbf{MEME Motif} & \textbf{Motif Logo} & \textbf{Top TF match(s)}  & Family\\ \hline
0 & DGTCCCWDGRGR & \includegraphics[height=0.8cm]{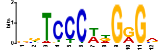} & Ebf2, EBF3, EBF1, Ebf4 & Early B-Cell Factor \\ \hline
0 & GCCAKCWGVTGG & \includegraphics[height=0.8cm]{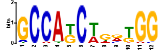} & CTCF, Atoh1, TCF21, Tcf12 &  C2H2 zinc finger,\\ 
& & & Twist2, BHLHE22, Neurod2 & bHLH \\ \hline
1 & HRMCCMCWGGGG & \includegraphics[height=0.8cm]{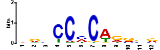} & 	
EBF1, Ebf4, Runx1, TFAP2A & Early B-Cell Factor, \\ 
& & & Ebf2, EBF3 & Runt\\ \hline
2 & RRCCACHDGSBS & \includegraphics[height=0.8cm]{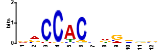} & TCF, Runx1, BCL11B & Runt, bHLH \\ 
& & & TCF21, BHLHE22, TFAP2A& \\ \hline
3 & YGCCMTCTWSTG & \includegraphics[height=0.8cm]{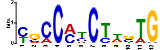} & CTCF & C2H2 zinc finger \\ \hline
4 & WRACCACARVVN & \includegraphics[height=0.8cm]{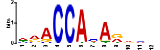} & Runx1, Runx2, Runx3 & Runt \\ \hline
4 & TTSCTGGT & \includegraphics[height=0.8cm]{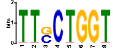} & IKZF1, ELF4, ELF2 , & Ets-related \\ 
& & & FOXO1::ELK3 & \\ \hline
5 & CYCTGTGGTBD & \includegraphics[height=0.8cm]{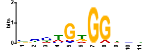} & Runx1, Bcl11B & Runt, C2H2 zinc finger \\ \hline
8  & CCACMAGATGGC & \includegraphics[height=0.8cm]{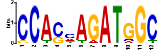} & CTCF, Atoh1, TCF21, Tcf12 &  C2H2 zinc finger,\\ 
& & & Twist2, BHLHE22, Neurod2 & bHLH \\ \hline
8 & NVBHTGTGGTYW & \includegraphics[height=0.8cm]{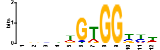} & Runx1, Runx2, Runx3 & Runt \\ \hline

\end{tabular}
\end{table}

These results indicate that WTKO-CNN not only provides accurate predictions of condition-specific accessibility but also enables mechanistic insights by linking sequence patterns to potential regulatory elements affected by gene knockout. By combining predictive performance with interpretability, the framework offers a powerful approach for dissecting the sequence basis of differential chromatin landscapes.

\subsection{Discriminative Motifs Identified by CNN Filters}

Saliency analysis identifies sequence positions that contribute most strongly to individual model predictions, whereas CNN filter-derived k-mers capture recurrent sequence patterns learned across the dataset. Together, these complementary approaches provide insights into both sequence-specific and shared regulatory features underlying model predictions. To characterize the sequence features learned by the model, we analyzed activations from the first convolutional layer (64 filters; kernel size = 10) across five independent training runs. For each filter, the highest-scoring activations were identified across all input sequences, and the corresponding DNA k-mers were extracted.

The extracted k-mers were aggregated to generate consensus motifs for each CNN filter, representing sequence patterns that maximally activate individual filters. These CNN-derived motifs were subsequently compared against known transcription factor binding motifs using \texttt{Tomtom} against the JASPAR 2022 vertebrate database, with significant matches defined as $p < 0.05$. The identified motifs were further characterized using motif clustering and MEME-based motif analysis.

Tables~\ref{tab:motifs},~\ref{tab:motifs1}, and~\ref{tab:motifs3} summarize the top discriminative motifs identified from WT versus KO RelA, LSD1, and Ebf1 peak regions. The CNN-derived motifs correspond to transcription factor families with established roles in transcriptional regulation, chromatin accessibility, and cellular signaling, suggesting that the model learns biologically meaningful sequence features associated with differential regulatory activity.

In the RelA dataset, CNN filters identified motifs corresponding to AP-1 family transcription factors, including FOS, JUN, and BATF. These findings are consistent with known functional interactions between AP-1 and NF-$\kappa$B signaling pathways. In addition, GC-rich motifs associated with SP family transcription factors (SP1, SP2, and SP4) were detected, suggesting that GC-binding regulatory elements may contribute to differences in chromatin and transcriptional regulation between WT and KO conditions. These results demonstrate that CNN-based motif discovery can recover regulatory signatures associated with condition-specific binding differences.

For the LSD1 dataset, MEME analysis of CNN-selected sequence clusters identified GC-rich motifs, including CCSCGCGCCC. Tomtom comparison matched these motifs to transcription factor families such as KLF16, SP1, SP2, HES1, TCF4, and ZIC1. These motif associations suggest that CNN filters capture regulatory elements potentially linked to LSD1-dependent chromatin regulation and transcriptional control.

For the Ebf1 dataset, highly activated CNN filter k-mers were compared with known transcription factor binding motifs. Several CNN-derived motifs showed similarity to established regulatory patterns, including Runt family motifs associated with RUNX1, RUNX2, and RUNX3, as well as EBF family motifs associated with EBF1, EBF2, EBF3, and EBF4. These results indicate that the CNN model captures biologically relevant sequence features within Ebf1 regulatory regions and identifies transcription factor motifs potentially contributing to differential regulatory activity between WT and KO conditions (Table~\ref{tab:motifs3}).



\begin{table}[h!]
\centering
\caption{Discriminative motifs identified from WT versus KO RelA peaks. Motifs are shown with representative sequence logos, consensus motifs, and matched transcription factor families.}
\label{tab:motifs}
\begin{tabular}{|c|c|c|c|c|}
\hline
\textbf{Dataset} & \textbf{Motif Logo} & \textbf{MEME MOTIF} &  \textbf{Matched TFs} & Family\\ \hline

RelA & \includegraphics[height=0.8cm]{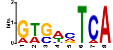} & GWSWCTCA & FOS, ATF3, BNC2 & bZIP\\ \hline
RelA & \includegraphics[height=0.8cm]{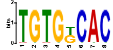} & TGTGKCAC & BATF::JUN, FOS::JUN, FOSB::JUNB & bZIP\\ \hline
RelA & \includegraphics[height=0.8cm]{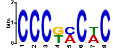} & CCCKMCWC & SP1,SP2, SP4 & C2H2\\ \hline

\end{tabular}
\end{table}
\begin{table}[h!]
\centering
\caption{Discriminative motifs identified from WT versus KO LSD1 peaks. Motifs are shown with representative sequence logos, consensus motifs, and matched transcription factor families.}
\label{tab:motifs1}
\begin{tabular}{|c|c|c|c|c|}
\hline
\textbf{Dataset} & \textbf{Motif Logo} & \textbf{MEME MOTIF} &  \textbf{Matched TFs} & Family \\ \hline
LSD1 & \includegraphics[height=0.8cm]{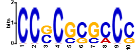} & 	CCSCGCGCCC & KLF16, SP1, SP2, KLF1, KLF12 & C2H2\\ \hline
LSD1 & \includegraphics[height=0.8cm]{lk.png} & CCSCGCGCCC & HES1, TCF4, ZIC1 & bHLH\\ \hline

\end{tabular}
\end{table}

\begin{table}[h!]
\centering
\caption{Discriminative motifs identified from WT versus KO Ebf1 peaks. Motifs are shown with representative sequence logos, consensus motifs, and matched transcription factor families.}
\label{tab:motifs3}
\begin{tabular}{|c|c|c|c|c|}
\hline
\textbf{Dataset} & \textbf{Motif Logo} & \textbf{MEME MOTIF} &  \textbf{Matched TFs} & Family\\ \hline

Ebf1 & \includegraphics[height=0.8cm]{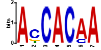} & ACCACAA & Runx1, Runx2, Runx3 & Runt\\ \hline
Ebf1 & \includegraphics[height=0.8cm]{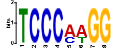} & TCCCMWGG & Ebf1, Ebf2, Ebf3, Ebf4 & EBF\\ \hline
Ebf1 & \includegraphics[height=0.8cm]{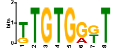} & TTGTGGKT & Runx1, Runx2, Runx3 & Runt\\ \hline

\end{tabular}
\end{table}
\section{Validation}
To validate the regulatory motifs identified by our WTKO-CNN model, we performed motif enrichment analysis using HOMER on high-saliency peaks from WT and KO test datasets (held out chromosomes). 
\subsection{HOMER Motif Enrichment Of RelA Chr6 and Chr7 WT and KO Peaks Validates WTKO-CNN Predictions}
RelA WT and KO test sequences derived from Chr6 and Chr7 are used for motif enrichment analysis. The sequences are analyzed using the HOMER motif discovery tool \cite{30} (findMotifsGenome.pl) to identify significantly enriched motifs at peak regions. HOMER performs enrichment analysis by comparing the input sequences against background genomic regions and identifying overrepresented motifs that may correspond to regulatory elements.
\begin{table}[h!]
\centering
\caption{HOMER motif enrichment analysis of RelA Chr6 and Chr7 peaks for WT and KO datasets. Top transcription factor motifs show differential prevalence between conditions.}
\label{tab:homer_enrichment}
\begin{tabular}{|l|l|c|c|c|c|}
\hline
\textbf{Motif Name} & \textbf{Consensus} & \textbf{\% Target (WT)} & \textbf{\% Target (KO)} & \textbf{p-value (WT)} & \textbf{p-value (KO)} \\
\hline
FOS (bZIP) & NDATGASTCAYN & 36.95\% & 59.80\% & $1\times10^{-87}$ & $1\times10^{-268}$ \\
ATF3 (bZIP) & DATGASTCATHN & 36.95\% & 60.46\% & $1\times10^{-79}$ & $1\times10^{-254}$ \\
BATF (bZIP) & DATGASTCAT & 37.47\% & 60.46\% & $1\times10^{-80}$ & $1\times10^{-252}$ \\
JunB (bZIP) & RATGASTCAT & 35.14\% & 56.86\% & $1\times10^{-81}$ & $1\times10^{-253}$ \\
Fosl2 (bZIP) & NATGASTCABNN & 26.87\% & 44.12\% & $1\times10^{-69}$ & $1\times10^{-218}$ \\
KLF5 (C2H2 ZF) & DGGGYGKGGC & 20.67 \% & 7.03\% & $1\times10^{-7}$ & $1\times10^{-4}$\\
MafK (bZIP) & GCTGASTCAGCA & 6.46\% & 12.58\% & $1\times10^{-4}$ & $1\times10^{-27}$ \\
\hline
\end{tabular}
\end{table}
Table~\ref{tab:homer_enrichment} summarizes the most enriched motifs, including the consensus sequence, percentage of target peaks containing the motif, and associated p-values. Notably, several bZIP family TFs, such as FOS, ATF3, BATF, JunB, and Fosl2, showed higher enrichment in KO peaks compared to WT, both in terms of the fraction of peaks containing the motif and statistical significance. In contrast, the C2H2 zinc finger TF KLF5 exhibited stronger enrichment in WT peaks, with a reduction in KO, suggesting differential regulation. These results highlight that loss of RelA alters the chromatin landscape, preferentially affecting motifs recognized by bZIP TFs, and provides a basis for linking motif changes to transcriptional reprogramming in KO cells. Figure 8 shows the percentage of target sequences containing each motif in WT and KO conditions. A full list of all enriched motifs and associated statistics is provided in Supplementary File S2 and S3.
\begin{figure}
    \centering
    \includegraphics[width=0.9\linewidth]{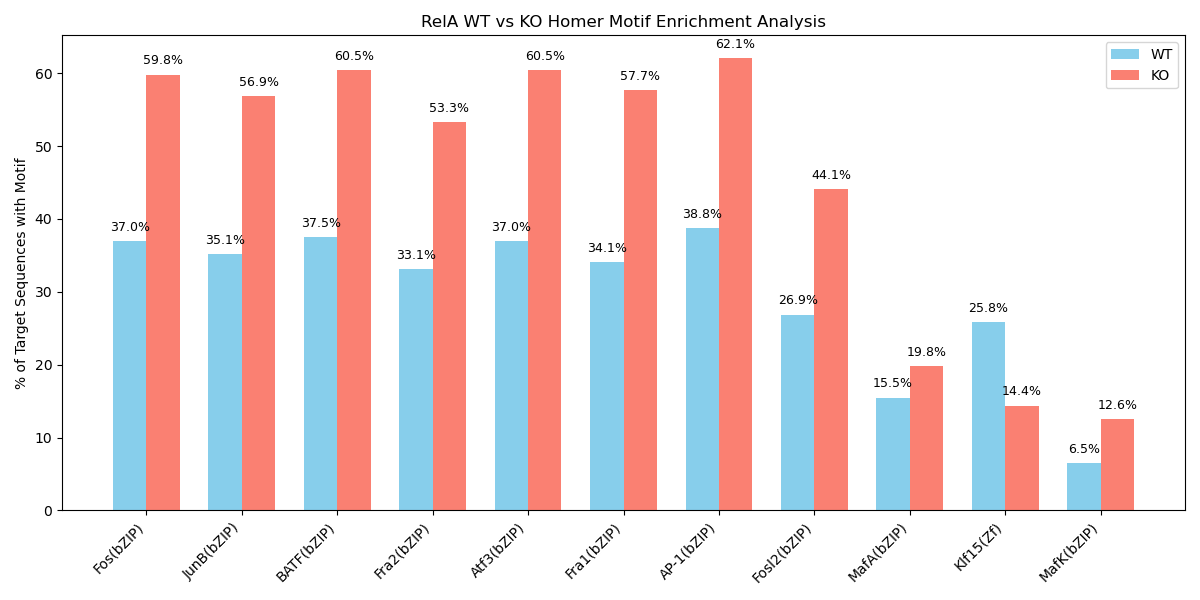}
    \caption{RelA WT vs KO Chr6 and Chr7 Homer Motif Enrichment Analysis.}
    \label{fig:homer}
\end{figure}
These results provide independent validation of the proposed model's predictions and demonstrate that the identified motifs reflect genuine transcription factor binding preferences. While HOMER identifies statistically enriched motifs in target sequences, it often generates long lists of motifs, including many redundant or non-discriminative hits. In contrast, our WTKO-CNN framework directly learns sequence features that drive classification between WT and KO conditions. By combining saliency mapping and clustering of top k-mers, we focus on biologically relevant motifs, capturing context-dependent and combinatorial patterns that HOMER alone may overlook.

\subsection{HOMER Motif Enrichment Of LSD1 Chr6 and Chr7 WT and KO Peaks Validates WTKO-CNN Predictions}
To investigate transcription factor (TF) binding changes upon LSD1 deletion, I analyzed both LSD1 WT and KO samples to assess changes in motif enrichment. Homer motif analysis revealed on whole LSD1 WT and KO that, bHLH transcription factor motifs are significantly enriched at peak regions, consistent with active binding of factors such as HEB, E2A, Ascl1, and Ptf1a (Supplenentary File 2). In contrast, in LSD1 knockout (KO) cells, bHLH motifs are largely absent, indicating a loss of accessibility at these sites. This suggests that bHLH factor binding is required for LSD1 chromatin states, and its deletion disrupts the recruitment of these transcriptional regulators. For LSD1 chromosomes 6 and 7, shared motifs highlighted conserved regulatory elements, with ETS family motifs, including ETS1 and ELF5 (Figure 9). bHLH family motifs, such as NeuroG2, Twist2, and BHLHA15, are also present in both conditions. Quantification of target sequences containing each motif revealed differences in relative enrichment; for example, NeuroG2 and BHLHA15 showed higher prevalence in WT, suggesting potential compensatory regulatory activity or altered chromatin accessibility. These trends are visualized using bar plots showing the percentage of target sequences with each motif, providing a clear view of transcription factor occupancy shifts associated with LSD1 loss on chromosomes 6 and 7.
\begin{table}[h!]
\centering
\caption{HOMER motif enrichment analysis of LSD1 Chr6 and Chr7 peaks for WT and KO datasets. Top transcription factor motifs show differential prevalence between conditions.}
\label{table:homer_results1}
\begin{tabular}{|l|l|c|c|c|c|}
\hline
\textbf{Motif Name} & \textbf{Consensus} & \textbf{\% Target (WT)} & \textbf{\% Target (KO)} & \textbf{p-value (WT)} & \textbf{p-value (KO)} \\
\hline
ETS1 (ETS) & ACAGGAAGTG & 44.68\% & 19.32\% & $1\times10^{-7}$ & $1\times10^{-1}$ \\
ELF5(ETS) & ACVAGGAAGT & 29.79\% & 12.50\% & $1\times10^{-5}$ & $1\times10^{-1}$ \\
Fli1(ETS) & NRYTTCCGGH & 36.17\% & 19.32\% & $1\times10^{-4}$ & $1\times10^{-1}$ \\
ERG(ETS) & ACAGGAAGTG & 44.68\% & 22.73\% & $1\times10^{-4}$ & $1\times10^{-1}$ \\
Twist2(bHLH) & MCAGCTGBYH & 51.06\% & 9.09\% & $1\times10^{-4}$ & $1\times10^{-0}$ \\
Max(bHLH)& 	RCCACGTGGYYN & 	25.5\% & 4.55\% & $1\times10^{-3}$ & $1\times10^{-0}$ \\
Ptf1a(bHLH)& ACAGCTGTTN & 	63.83\% & 22.73 \% & $1\times10^{-3}$ & $1\times10^{-0}$ \\

\hline
\end{tabular}
\end{table}
\begin{figure}
    \centering
    \includegraphics[width=0.9\linewidth]{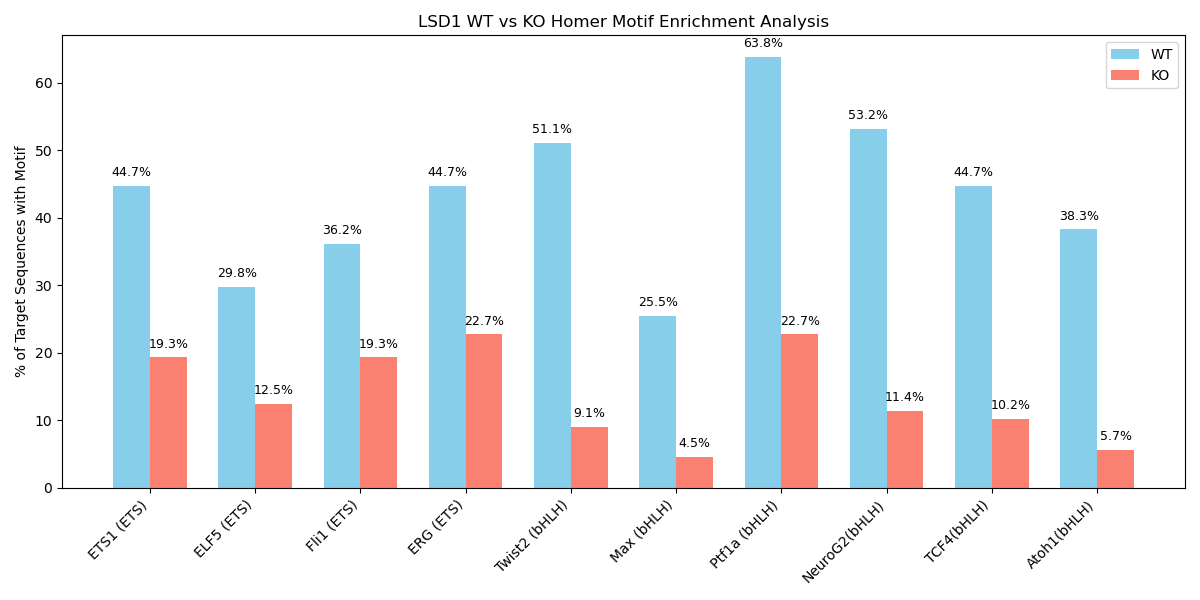}
    \caption{LSD1 WT vs KO Chr6 and Chr7 Homer Motif Enrichment Analysis.}
    \label{fig:homer2}
\end{figure}
For LSD1, Homer analysis of chromosomes 6 and 7 test sequences revealed that ETS motifs are more abundant in WT sequences. However, the CNN-attention model identified bHLH motifs as the primary discriminative TF. This indicates that, although ETS motifs are enriched in WT, the predictive power of the model for distinguishing WT from KO sequences relies predominantly on bHLH motifs, highlighting their functional importance in driving WT-specific chromatin accessibility.
\subsection{HOMER Motif Enrichment Of Ebf1 Chr6 and Chr7 WT and KO Peaks Validates WTKO-CNN Predictions}
Motif enrichment analysis of WT-specific and EBF1KO-specific accessible regions revealed distinct transcription factor regulatory signatures. WT-specific peaks showed strong enrichment for CTCF/BORIS zinc-finger motifs, ETS family motifs, and EBF family motifs, including EBF1, EBF2, ETS1, PU.1, and related ETS transcription factors (Supplenentary File 2 and 3). The presence of EBF1 and ETS-family motifs in WT-specific regions is consistent with the establishment of B-cell lineage-associated regulatory elements and suggests that EBF1-dependent chromatin accessibility is associated with B-cell transcriptional programs. In contrast, EBF1KO-specific regions were highly enriched for RUNX family motifs, particularly RUNX1/RUNX-related binding sites, along with ETS-family and AP-1/bZIP motifs. The enrichment of RUNX and AP-1 regulatory elements in knockout-specific regions suggests activation or exposure of alternative regulatory programs following loss of EBF1. Together, these results indicate that EBF1-dependent chromatin regions are associated with B-cell identity factors, whereas EBF1 loss leads to increased accessibility of regions enriched for transcription factors linked to alternative lineage regulation and cellular plasticity.
\begin{table}[h!]
\centering
\caption{HOMER motif enrichment analysis of Ebf1 Chr6 and Chr7 peaks for WT and KO datasets. Top transcription factor motifs show differential prevalence between conditions.}
\label{table:homer_results1}
\begin{tabular}{|c|c|p{2 cm}|p{2 cm}|c|c|}
\hline
\textbf{Motif Name} & \textbf{Consensus} & \textbf{\% Target (WT)} & \textbf{\% Target (KO)} & \textbf{p-value (WT)} & \textbf{p-value (KO)} \\
\hline
CTCF (Zf) & AYAGTGCCMY &  14.82\% & 4.70\% & $1\times10^{-475}$ & $1\times10^{-23}$\\
& CTRGTGGCC & & & & \\
ERG (ETS) & ACAGGAAGTG & 35.19\% & 31.31\% & $1\times10^{-184}$ & $1\times10^{-66}$ \\
ETS1 (ETS) & ACAGGAAGTG & 26.19\% & 23.78\% & $1\times10^{-177}$ & $1\times10^{-74}$ \\
EBF (EBF) & DGTCCCYRGGGA & 8.75\% & 1.08\% & $1\times10^{-131}$ & $1\times10^{-0}$ \\
EBF1 (EBF) & GTCCCCWGGGGA & 21.02\% & 7.93\% & $1\times10^{-75}$ & $1\times10^{-0}$ \\
EBF2 (EBF) & NABTCCCWDGGGAVH & 22.81\% & 5.33\% & $1\times10^{-156}$ & $1\times10^{-0}$ \\

RUNX1 (Runt) & AAACCACARM & 14.63\% & 44.23\% & $1\times10^{-19}$ & $1\times10^{-264}$ \\
RUNX2 (Runt) & NWAACCACADNN & 11.54\%  & 38.85\% & $1\times10^{-15}$ & $1\times10^{-224}$ \\
RUNX (Runt) & SAAACCACAG & 10.33\%  & 34.05\% & $1\times10^{-18}$ & $1\times10^{-233}$ \\
NeuroD1 (bHLH) & GCCATCTGTT & 15.07\%  & 11.04\% & $1\times10^{-42}$ & $1\times10^{-2}$ \\
Tcf12 (bHLH) & NAACAGCTGG & 19.41\%  & 11.59\% & $1\times10^{-52}$ & $1\times10^{-1}$ \\
 
\hline
\end{tabular}
\end{table}
\begin{figure}
    \centering
    \includegraphics[width=0.9\linewidth]{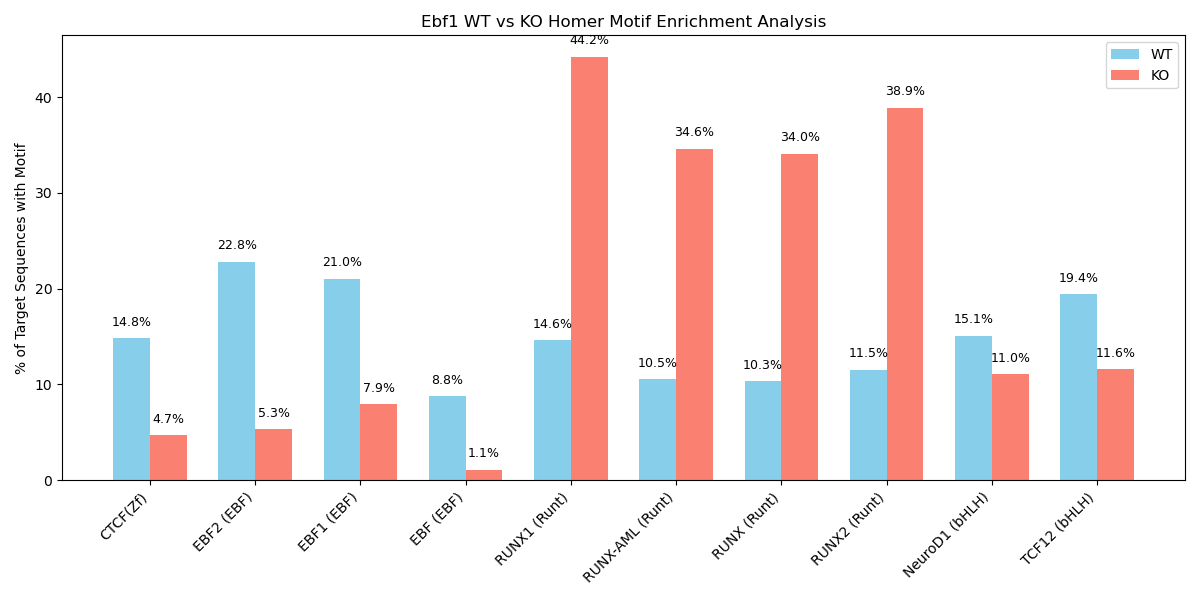}
    \caption{Ebf1 WT vs KO Chr6 and Chr7 Homer Motif Enrichment Analysis.}
    \label{fig:homer2}
\end{figure}

 HOMER motif analysis demonstrated that WT-specific peaks were enriched for EBF1, EBF2, ETS1, PU.1, and CTCF motifs, highlighting the association of EBF1-dependent accessibility with B-cell lineage transcriptional regulation. Conversely, EBF1KO-specific peaks showed strong enrichment for RUNX1/RUNX motifs, ETS motifs, and AP-1/bZIP motifs, suggesting increased accessibility of regulatory elements associated with alternative transcriptional programs after EBF1 loss.
\section{Comparison with Other Algorithms}
I examined several sequence classification models: CNN with 4 Layers, a hybrid CNN+LSTM model and Vanilla Transformer model to assess how well various deep learning architectures predicted chromatin accessibility patterns between WT and KO situations.
The proposed WTKO-CNN using sequence input performed the best out of all the models , demonstrating its great ability to extract and use local sequence characteristics that are essential for differentiating between WT and KO peaks. 

I implemented a deep learning architecture consisting of four stacked convolutional neural network (CNN) layers. This model is designed to progressively extract hierarchical features from DNA sequences, where early layers capture simple local motifs and deeper layers learn more complex and abstract sequence patterns. By stacking multiple convolutional layers, the network is able to effectively model spatial dependencies within short genomic regions that are often biologically relevant.

In the CNN+LSTM model, a Long Short-Term Memory (LSTM) network is applied after the CNN layers to capture long-range dependencies and sequential patterns in the data. This hybrid architecture enables the model to learn relationships between nucleotides that may be distant in the sequence, but biologically significant. The LSTM layer retains important contextual information throughout the sequence, complementing the local feature extraction performed by the CNN. Finally, the output of the LSTM is passed through a fully connected layer for classification, allowing the model to predict whether a given DNA sequence belongs to the WT or KO category. This combined approach leverages the strengths of CNNs in capturing local motifs and LSTMs in modeling global sequence dependencies, making it particularly effective for DNA sequence classification tasks where both short- and long-range interactions are important.

I also implemented a Transformer-based model to predict chromatin accessibility states for WT and KO sequences. Unlike BERT-based approaches that rely on large-scale pretraining, our vanilla Transformer is trained from scratch on the target dataset. The model begins with an embedding layer that converts DNA sequences into dense numerical representations. These embeddings are then passed through multiple Transformer encoder layers, which use self-attention mechanisms to capture long-range dependencies and complex interactions across the sequence. This allows the model to learn global contextual relationships without relying on recurrence or convolution. The final representation is fed into a classification head to predict chromatin accessibility states.

The CNN+Transformer model demonstrated comparable performance across the three datasets. For GSE107075 (RelA), the model achieved a precision of 0.70, recall of 0.40, and F1-score of 0.51 for the WT class, while the KO class achieved a precision of 0.58, recall of 0.82, and F1-score of 0.68, resulting in an overall accuracy of 62

For GSE119222 (LSD1), the model showed similar performance, with WT achieving precision, recall, and F1-score values of 0.55, 0.71, and 0.62, respectively, and KO achieving values of 0.73, 0.57, and 0.64, respectively. The model obtained an overall accuracy of 63

For GSE295902 (Ebf1), the CNN+Transformer model achieved a WT precision of 0.66, recall of 0.85, and F1-score of 0.74, while the KO class achieved a precision of 0.81, recall of 0.60, and F1-score of 0.69, resulting in an overall accuracy of 72\%. Overall, the CNN+Transformer model maintained balanced performance across different regulatory datasets, capturing sequence patterns associated with WT and KO conditions.

The Table 17 shows that the proposed WTKO-CNN model consistently outperforms the other evaluated architectures, including CNN, CNN+LSTM, CNN+Transfomer and the vanilla Transformer, across three datasets. It achieves the highest overall accuracy, reaching up to 84\% on LSD1, demonstrating a substantial improvement over competing methods. In addition to improved accuracy, WTKO-CNN shows stronger and more balanced performance in precision, recall, and F1-score for both WT and KO classes, indicating better discrimination capability and reduced classification bias (Figure 11).
Average F1-score comparison of different deep learning models across RelA, LSD1, and Ebf1 datasets are also mention in Figure 11(e). 
This improvement can be attributed to the model’s tailored architecture, which is specifically designed to capture class-relevant patterns distinguishing WT and KO sequences. Unlike CNN-only models that primarily focus on local motif detection or CNN+LSTM models that may struggle with long-range dependency optimization, WTKO-CNN effectively integrates feature learning in a way that enhances both local and global sequence representation. Compared to the Transformer model trained from scratch, WTKO-CNN demonstrates better generalization on limited genomic datasets, likely due to its more efficient inductive bias for sequence-based biological signals. Overall, these results highlight the robustness and effectiveness of WTKO-CNN for chromatin accessibility and DNA sequence classification tasks.
\begin{table}[ht]
\centering
\caption{Performance comparison of different models across datasets.}
\small
\begin{tabular}{llccccc}
\toprule
Algorithm & Dataset & Class & Precision & Recall & F1-score & Accuracy \\
\midrule

\multirow{6}{*}{CNN 4 layers}
& \multirow{2}{*}{GSE107075 (RelA)} & WT & 0.69 & 0.49 & 0.57 & \multirow{2}{*}{64\%} \\
& & KO & 0.61 & 0.79 & 0.68 & \\
\cmidrule(lr){2-7}
& \multirow{2}{*}{GSE119222 (LSD1)} & WT & 0.42 & 0.59 & 0.49 & \multirow{2}{*}{65\%} \\
& & KO & 0.65 & 0.89 & 0.75 & \\
\cmidrule(lr){2-7}
& \multirow{2}{*}{GSE295902 (Ebf1)} & WT & 0.72 & 0.86 & 0.79 & \multirow{2}{*}{73\%} \\
& & KO & 0.74 & 0.56 & 0.64 & \\
\midrule

\multirow{6}{*}{CNN+LSTM}
& \multirow{2}{*}{GSE119222 (RelA)} & WT & 0.65 & 0.54 & 0.59 & \multirow{2}{*}{63\%} \\
& & KO & 0.61 & 0.71 & 0.66 & \\
\cmidrule(lr){2-7}
& \multirow{2}{*}{GSE107075 (LSD1)} & WT & 0.59 & 0.68 & 0.63 & \multirow{2}{*}{54\%} \\
& & KO & 0.44 & 0.35 & 0.39 & \\
\cmidrule(lr){2-7}
& \multirow{2}{*}{GSE295902 (Ebf1)} & WT & 0.78 & 0.76 & 0.77 & \multirow{2}{*} {78\%}\\
& & KO & 0.78 & 0.81 & 0.80 &\\
\midrule

\multirow{6}{*}{Vanilla Transformer}
& \multirow{2}{*}{GSE107075 (RelA)} & WT & 0.58 & 0.77 & 0.66 & \multirow{2}{*}{60\%} \\
& & KO & 0.65 & 0.44 & 0.53 & \\
\cmidrule(lr){2-7}
& \multirow{2}{*}{GSE119222 (LSD1)} & WT & 0.59 & 0.75 & 0.66 & \multirow{2}{*}{62\%} \\
& & KO & 0.68 & 0.50 & 0.51 & \\
\cmidrule(lr){2-7}
& \multirow{2}{*}{GSE295902 (Ebf1)} & WT & 0.59 & 0.65 & 0.62 & \multirow{2}{*} {67\%}\\
& & KO & 0.72 & 0.66 & 0.69 &\\
\midrule
\multirow{6}{*}{CNN+Transformer}
& \multirow{2}{*}{GSE107075 (RelA)} & WT & 0.70 & 0.40 & 0.51 & \multirow{2}{*}{62\%} \\
& & KO & 0.58 & 0.82 & 0.68 & \\
\cmidrule(lr){2-7}
& \multirow{2}{*}{GSE119222 (LSD1)} & WT & 0.55 & 0.71 & 0.62 & \multirow{2}{*}{63\%} \\
& & KO & 0.73 & 0.57 & 0.64 & \\
\cmidrule(lr){2-7}
& \multirow{2}{*}{GSE295902 (Ebf1)} & WT & 0.66 & 0.85 & 0.74 & \multirow{2}{*} {72\%}\\
& & KO & 0.81 & 0.60 & 0.69 &\\
\midrule
\multirow{6}{*}{\textbf{WTKO-CNN}}
& \multirow{2}{*}{GSE107075 (RelA)} & WT & 0.73 & 0.57 & 0.64 & \multirow{2}{*}{\textbf{68\%}} \\
& & KO & 0.65 & 0.79 & 0.71 & \\
\cmidrule(lr){2-7}
& \multirow{2}{*}{GSE119222 (LSD1)} & WT & 0.83 & 0.91 & 0.87 & \multirow{2}{*} {\textbf{84\%}} \\
& & KO & 0.87 & 0.74 & 0.80 & \\
\cmidrule(lr){2-7}
& \multirow{2}{*}{GSE295902 (Ebf1)} & WT & 0.78 & 0.79 & 0.78 & \multirow{2}{*} {\textbf{79\%}} \\
& & KO & 0.81 & 0.79 & 0.80 &\\
\bottomrule
\end{tabular}
\label{tab:results}
\end{table}

\begin{figure}[htbp]
    \centering

    \begin{subfigure}[b]{0.45\textwidth}
        \centering
        \includegraphics[width=\textwidth]{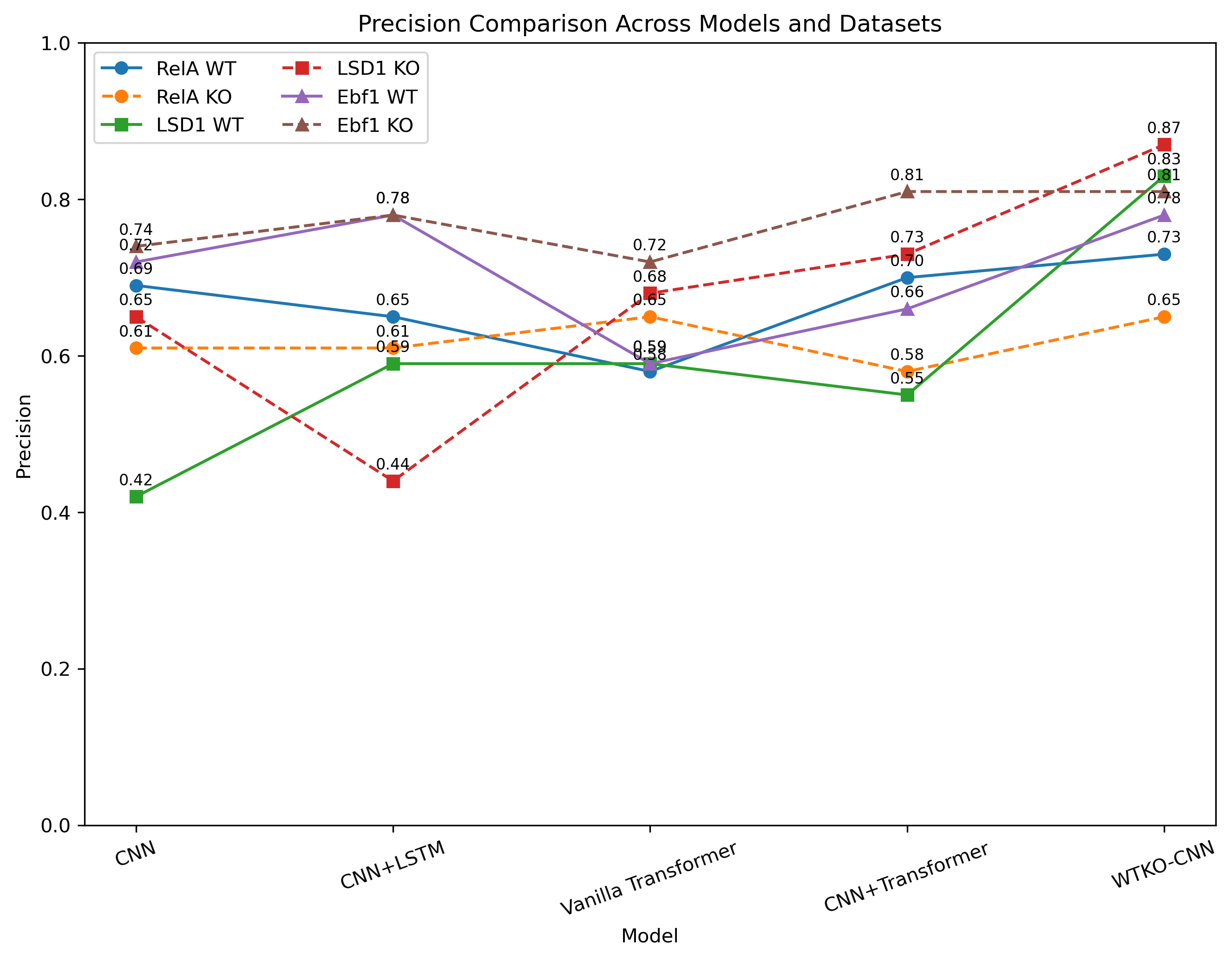}
        \caption{Precision}
        \label{fig:precision}
    \end{subfigure}
    \begin{subfigure}[b]{0.45\textwidth}
        \centering
        \includegraphics[width=\textwidth]{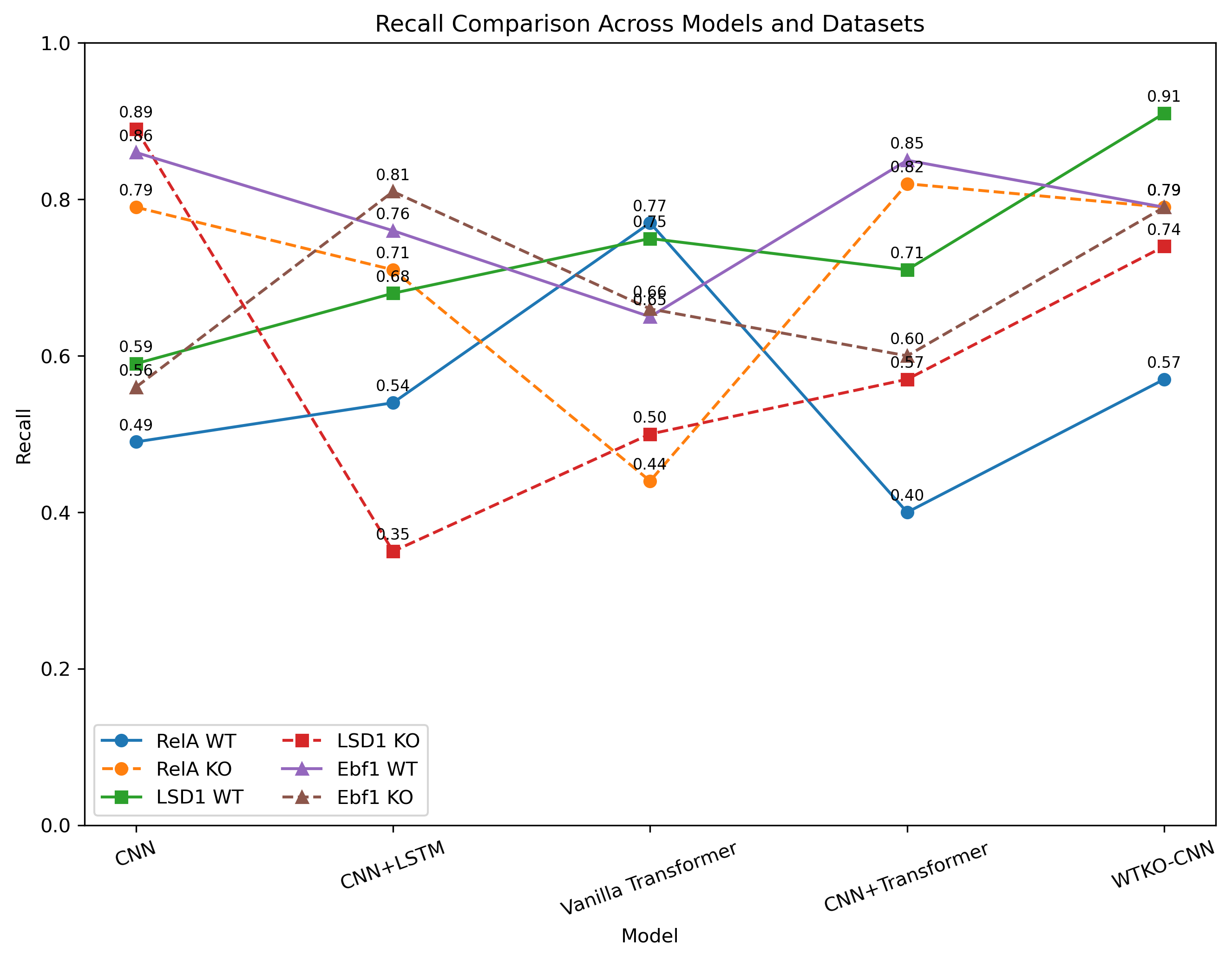}
        \caption{Recall}
        \label{fig:recall}
    \end{subfigure}
       \begin{subfigure}[b]{0.45\textwidth}
        \centering
        \includegraphics[width=\linewidth]{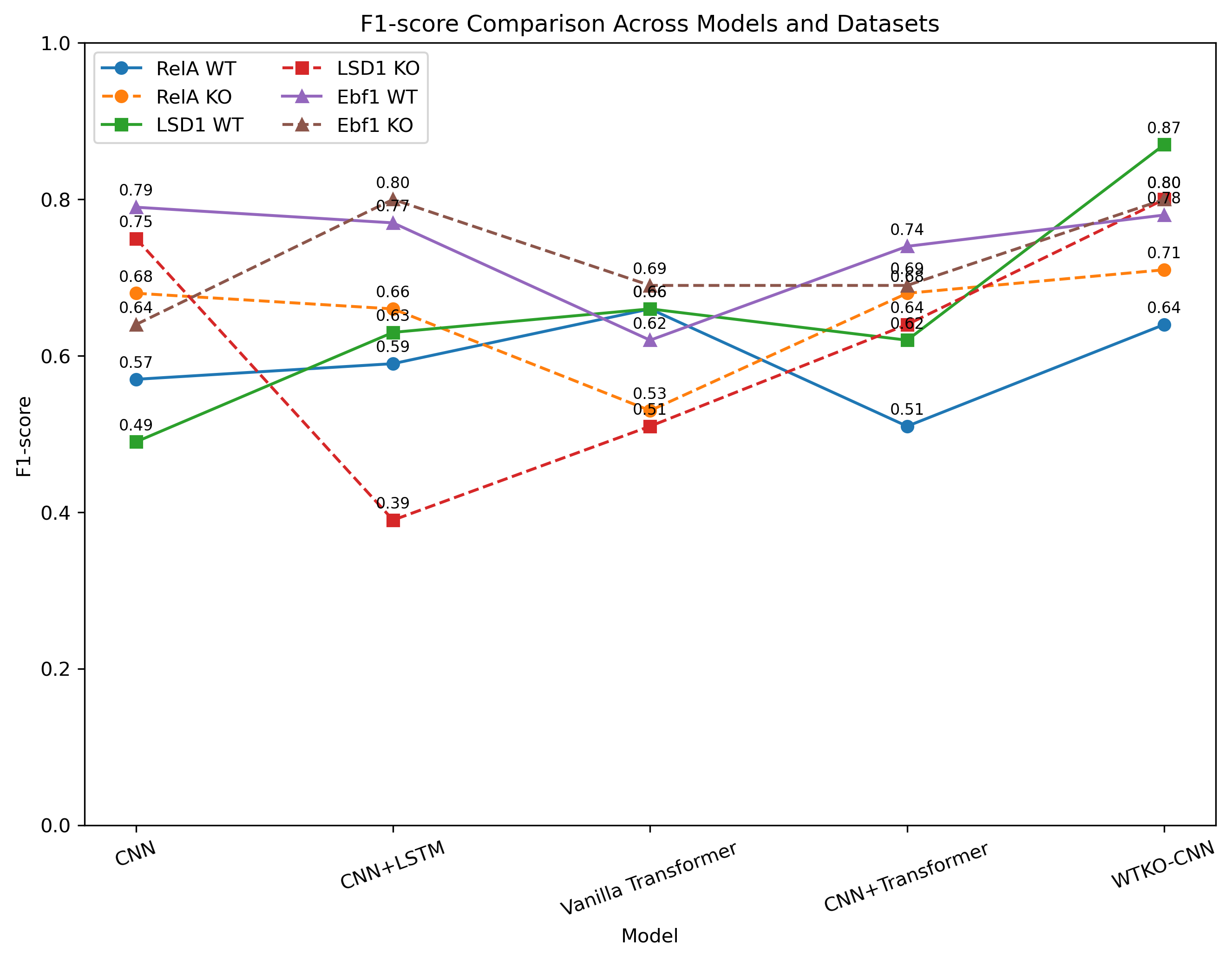}
        \caption{F1-Score}
        \label{fig:sub1}
    \end{subfigure}
  \begin{subfigure}[b]{0.45\textwidth}
        \centering
        \includegraphics[width=\linewidth]{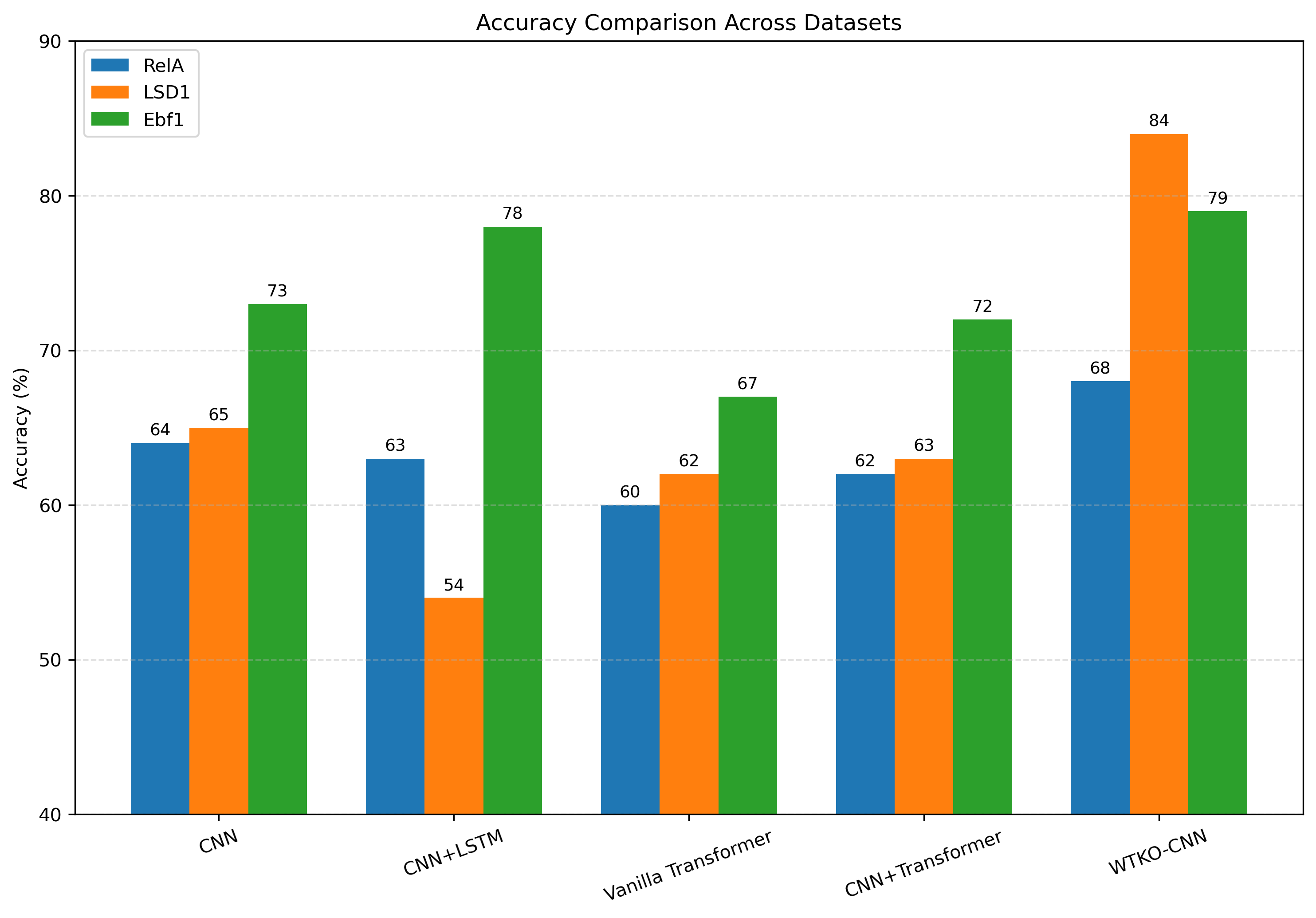}
        \caption{Accuracy}
        \label{fig:sub1}
    \end{subfigure}
    \begin{subfigure}[b]{0.5\textwidth}
        \centering
        \includegraphics[width=\linewidth]{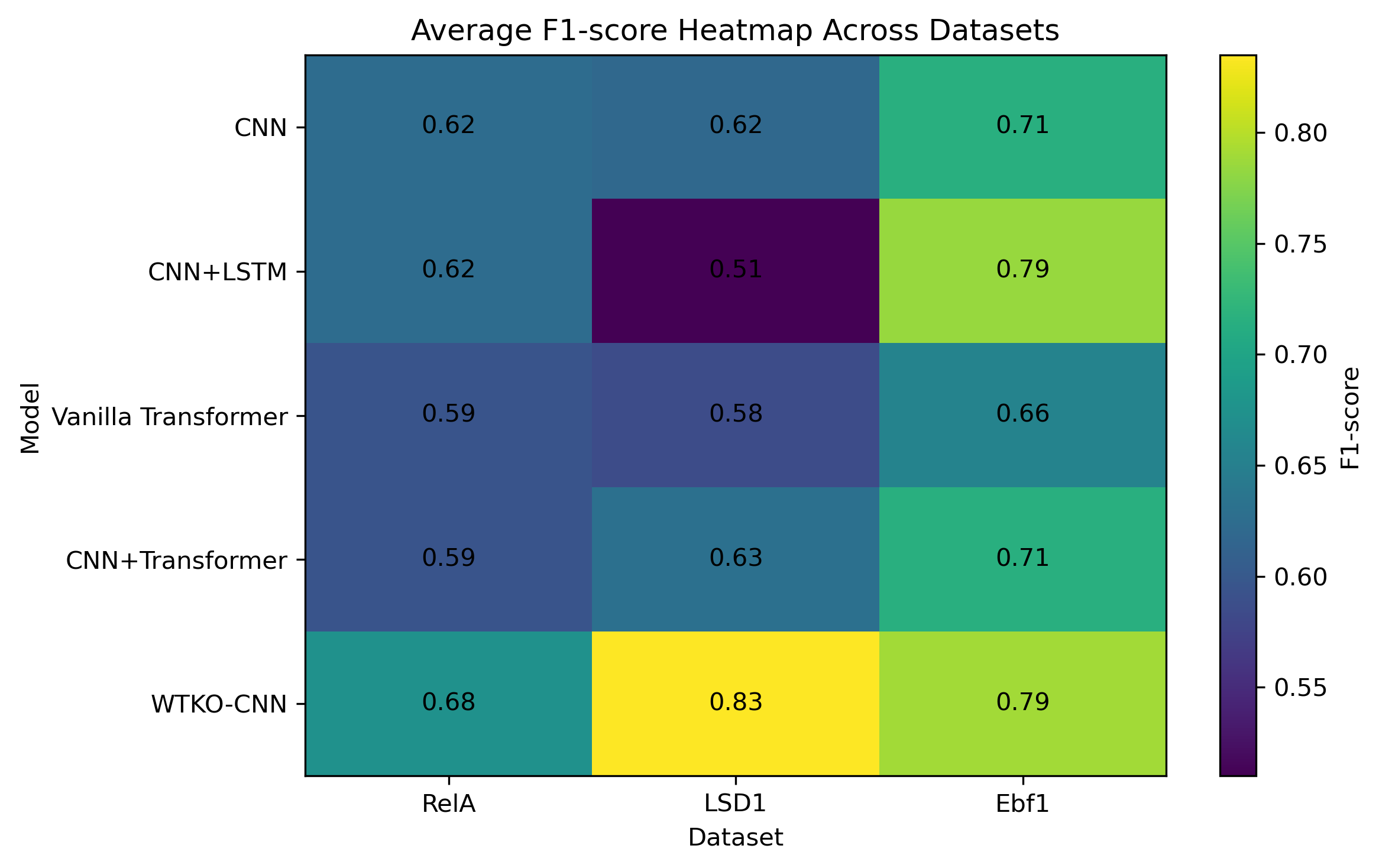}
        \caption{Average F1 Score}
        \label{fig:sub1}
    \end{subfigure}
    \caption{Comparative evaluation of model performance across RelA, LSD1, and Ebf1 datasets. The subfigures show (a) Precision, (b) Recall, (c) F1-score, (d) Accuracy, and (e) average F1-score comparisons among different deep learning architectures. The proposed WTKO-CNN model demonstrates improved classification performance compared with baseline approaches, showing enhanced discrimination between WT and KO classes across multiple regulatory datasets.}

    \label{fig:all_metrics}
\end{figure}
\section{Conclusion}

Understanding transcriptional regulatory mechanisms requires identifying the sequence features that drive condition-specific chromatin accessibility. To address this challenge, I developed WTKO-CNN, a one-dimensional convolutional neural network incorporating an attention mechanism to classify genomic sequences as either WT- or KO-specific. While conventional motif-enrichment approaches such as HOMER successfully identified primary overrepresented motifs associated with known transcription factor families, the CNN-attention framework provided additional insight into the regulatory sequence features underlying differential accessibility.

Unlike frequency-based motif enrichment approaches, saliency map analysis identified nucleotide regions that contributed most strongly to model predictions, enabling the discovery of biologically relevant sequence patterns beyond the most abundant motifs. This approach revealed both canonical and non-canonical motifs, as well as higher-order sequence features that may reflect secondary transcription factor recruitment, cooperative binding interactions, or DNA sequence constraints associated with altered chromatin accessibility in KO conditions.

Across the RelA, LSD1, and Ebf1 datasets, WTKO-CNN consistently outperformed conventional CNN and hybrid deep learning architectures, achieving the highest classification accuracy for all evaluated datasets. In particular, WTKO-CNN achieved accuracies of 68\%, 84\%, and 79\% for RelA, LSD1, and Ebf1, respectively, demonstrating its ability to capture dataset-specific regulatory patterns. Furthermore, CNN-derived motifs identified through filter activation and motif analysis showed correspondence with known transcription factor families, including AP-1, SP, KLF, Runt, and EBF-associated motifs, supporting the biological relevance of the learned sequence representations.

Overall, WTKO-CNN provides an interpretable deep learning framework that integrates sequence classification with motif discovery to characterize regulatory elements associated with condition-specific chromatin accessibility. By combining predictive performance with model interpretability, this approach offers a complementary strategy to traditional motif enrichment methods and provides new opportunities for studying sequence determinants of transcriptional regulation. 

In future, I am planning to  integrate additional genomic and epigenomic information, such as histone modifications, DNA methylation, and transcription factor binding data, to improve the identification of regulatory mechanisms. Furthermore, expanding WTKO-CNN to multi-class and multi-condition analyses would enable the investigation of more complex regulatory landscapes across diverse cellular states and disease conditions.
\section*{Data and Code Availability}

The Data and source code for WTKO-CNN is available at:
\url{https://github.com/LopamudraDey/WTKO-CNN}.
The repository contains all datasets and scripts required for model training, evaluation, saliency-based k-mer extraction, and motif analysis. Instructions for reproducing the reported results are provided in the accompanying documentation.

\section*{Supplementary Information}

Supplementary data associated with this article are available online. \\
Supplementary File S1: S1 contains experimentally generated WT and KO peaks of RelA, LSD1 and Ebf1.\\
Supplementary Files S2 and S3: S2 contains Homer Motif Enrichment Results of Test Sequences, Chromosome 6 and 7 of RelA, LSD1 and Ebf1, and S3 contains the Homer Motif Enrichment Results of all Peaks of RelA, LSD1 and Ebf1.\\
Supplementary Folder S4: All TOMTOM results of each cluster MEME motifs. \\

\printbibliography

@article{4,
  title={Chromatin accessibility and transcription factor binding through the perspective of mitosis},
  author={Coux, R{\'e}mi-Xavier and Owens, Nick DL and Navarro, Pablo},
  journal={Transcription},
  volume={11},
  number={5},
  pages={236--240},
  year={2020},
  publisher={Taylor \& Francis}
}

@article{2,
  title={Transcription factors: from enhancer binding to developmental control},
  author={Spitz, Fran{\c{c}}ois and Furlong, Eileen EM},
  journal={Nature reviews genetics},
  volume={13},
  number={9},
  pages={613--626},
  year={2012},
  publisher={Nature Publishing Group UK London}
}

@article{3,
  title={MEDEA: analysis of transcription factor binding motifs in accessible chromatin},
  author={Mariani, Luca and Weinand, Kathryn and Gisselbrecht, Stephen S and Bulyk, Martha L},
  journal={Genome Research},
  volume={30},
  number={5},
  pages={736--748},
  year={2020},
  publisher={Cold Spring Harbor Lab}
}

@article{1,
  title={Early B-Cell Factor 1: An Archetype for a Lineage-Restricted Transcription Factor Linking Development to Disease},
  author={Sigvardsson, Mikael},
  journal={Transcription factors in blood cell development},
  pages={143--156},
  year={2024},
  publisher={Springer}
}

@article{5,
  title={ATAC-seq normalization method can significantly affect differential accessibility analysis and interpretation},
  author={Reske, Jake J and Wilson, Mike R and Chandler, Ronald L},
  journal={Epigenetics \& chromatin},
  volume={13},
  pages={1--17},
  year={2020},
  publisher={Springer}
}

@article{6,
  title={TRAFICA: Improving Transcription Factor Binding Affinity Prediction using Deep Language Model on ATAC-seq Data},
  author={Xu, Y and Wang, C and Xu, K and Ding, Y and Lyu, A and Zhang, L},
  year={2023}
}

@article{8,
  title={Convolutional neural network for automated peak detection in reversed-phase liquid chromatography},
  author={Kensert, Alexander and Bosten, Emery and Collaerts, Gilles and Efthymiadis, Kyriakos and Van Broeck, Peter and Desmet, Gert and Cabooter, Deirdre},
  journal={Journal of Chromatography A},
  volume={1672},
  pages={463005},
  year={2022},
  publisher={Elsevier}
}

@article{9,
  title={CNN-Peaks: ChIP-Seq peak detection pipeline using convolutional neural networks that imitate human visual inspection},
  author={Oh, Dongpin and Strattan, J Seth and Hur, Junho K and Bento, Jos{\'e} and Urban, Alexander Eckehart and Song, Giltae and Cherry, J Michael},
  journal={Scientific reports},
  volume={10},
  number={1},
  pages={7933},
  year={2020},
  publisher={Nature Publishing Group UK London}
}

@article{10,
  title={Peak-CNN: improved particle image localization using single-stage CNNs},
  author={Godbersen, Philipp and Schanz, Daniel and Schr{\"o}der, Andreas},
  journal={Experiments in Fluids},
  volume={65},
  number={10},
  pages={153},
  year={2024},
  publisher={Springer}
}

@article{11,
  title={A hybrid CNN-LSTM model for pre-miRNA classification},
  author={Tasdelen, Abdulkadir and Sen, Baha},
  journal={Scientific reports},
  volume={11},
  number={1},
  pages={14125},
  year={2021},
  publisher={Nature Publishing Group UK London}
}

@article{12,
  title={Analysis of DNA sequence classification using CNN and hybrid models},
  author={Gunasekaran, Hemalatha and Ramalakshmi, Krishnasamy and Rex Macedo Arokiaraj, A and Deepa Kanmani, S and Venkatesan, Chandran and Suresh Gnana Dhas, C},
  journal={Computational and Mathematical Methods in Medicine},
  volume={2021},
  number={1},
  pages={1835056},
  year={2021},
  publisher={Wiley Online Library}
}

@article{13,
  title={Comparing machine learning algorithms with or without feature extraction for DNA classification},
  author={Zhang, Xiangxie and Beinke, Ben and Kindhi, Berlian Al and Wiering, Marco},
  journal={arXiv preprint arXiv:2011.00485},
  year={2020}
}

@inproceedings{14,
  title={Statistical linear models in virus genomic alignment-free classification: application to hepatitis C viruses},
  author={Remita, Amine M and Diallo, Abdoulaye Banir{\'e}},
  booktitle={2019 IEEE International Conference on Bioinformatics and Biomedicine (BIBM)},
  pages={474--481},
  year={2019},
  organization={IEEE}
}

@article{17,
  title={Genetic compensation triggered by mutant mRNA degradation},
  author={El-Brolosy, Mohamed A and Kontarakis, Zacharias and Rossi, Andrea and Kuenne, Carsten and G{\"u}nther, Stefan and Fukuda, Nana and Kikhi, Khrievono and Boezio, Giulia LM and Takacs, Carter M and Lai, Shih-Lei and others},
  journal={Nature},
  volume={568},
  number={7751},
  pages={193--197},
  year={2019},
  publisher={Nature Publishing Group UK London}
}

@article{18,
  title={CAE-CNN: Predicting transcription factor binding site with convolutional autoencoder and convolutional neural network},
  author={Zhang, Yongqing and Qiao, Shaojie and Zeng, Yuanqi and Gao, Dongrui and Han, Nan and Zhou, Jiliu},
  journal={Expert Systems with Applications},
  volume={183},
  pages={115404},
  year={2021},
  publisher={Elsevier}
}

@inproceedings{19,
  title={Deep motif dashboard: visualizing and understanding genomic sequences using deep neural networks},
  author={Lanchantin, Jack and Singh, Ritambhara and Wang, Beilun and Qi, Yanjun},
  booktitle={Pacific symposium on biocomputing 2017},
  pages={254--265},
  year={2017},
  organization={World Scientific}
}

@article{20,
  title={Improving representations of genomic sequence motifs in convolutional networks with exponential activations},
  author={Koo, Peter K and Ploenzke, Matt},
  journal={Nature machine intelligence},
  volume={3},
  number={3},
  pages={258--266},
  year={2021},
  publisher={Nature Publishing Group UK London}
}

@article{22,
  title={DeFine: deep convolutional neural networks accurately quantify intensities of transcription factor-DNA binding and facilitate evaluation of functional non-coding variants},
  author={Wang, Meng and Tai, Cheng and E, Weinan and Wei, Liping},
  journal={Nucleic acids research},
  volume={46},
  number={11},
  pages={e69--e69},
  year={2018},
  publisher={Oxford University Press}
}

@article{23,
  title={Prediction of regulatory motifs from human Chip-sequencing data using a deep learning framework},
  author={Yang, Jinyu and Ma, Anjun and Hoppe, Adam D and Wang, Cankun and Li, Yang and Zhang, Chi and Wang, Yan and Liu, Bingqiang and Ma, Qin},
  journal={Nucleic acids research},
  volume={47},
  number={15},
  pages={7809--7824},
  year={2019},
  publisher={Oxford University Press}
}

@article{24,
  title={Beyond saliency: understanding convolutional neural networks from saliency prediction on layer-wise relevance propagation},
  author={Li, Heyi and Tian, Yunke and Mueller, Klaus and Chen, Xin},
  journal={Image and Vision Computing},
  volume={83},
  pages={70--86},
  year={2019},
  publisher={Elsevier}
}

@article{25,
  title={Deep-LIFT: Deep label-specific feature learning for image annotation},
  author={Li, Junbing and Zhang, Changqing and Zhou, Joey Tianyi and Fu, Huazhu and Xia, Shuyin and Hu, Qinghua},
  journal={IEEE transactions on Cybernetics},
  volume={52},
  number={8},
  pages={7732--7741},
  year={2021},
  publisher={IEEE}
}

@article{26,
  title={ExplaiNN: interpretable and transparent neural networks for genomics},
  author={Novakovsky, Gherman and Fornes, Oriol and Saraswat, Manu and Mostafavi, Sara and Wasserman, Wyeth W},
  journal={Genome Biology},
  volume={24},
  number={1},
  pages={154},
  year={2023},
  publisher={Springer}
}

@article{27,
  title={Predicting effects of noncoding variants with deep learning--based sequence model},
  author={Zhou, Jian and Troyanskaya, Olga G},
  journal={Nature methods},
  volume={12},
  number={10},
  pages={931--934},
  year={2015},
  publisher={Nature Publishing Group US New York}
}

@article{28,
  title={Histone demethylase LSD1 is required for germinal center formation and BCL6-driven lymphomagenesis},
  author={Hatzi, Katerina and Geng, Huimin and Doane, Ashley S and Meydan, Cem and LaRiviere, Reed and Cardenas, Mariano and Duy, Cihangir and Shen, Hao and Vidal, Maria Nieves Calvo and Baslan, Timour and others},
  journal={Nature immunology},
  volume={20},
  number={1},
  pages={86--96},
  year={2019},
  publisher={Nature Publishing Group US New York}
}

@article{29,
  title={Assessing deep learning methods in cis-regulatory motif finding based on genomic sequencing data},
  author={Zhang, Shuangquan and Ma, Anjun and Zhao, Jing and Xu, Dong and Ma, Qin and Wang, Yan},
  journal={Briefings in Bioinformatics},
  volume={23},
  number={1},
  pages={bbab374},
  year={2022},
  publisher={Oxford University Press}
}

@article{30,
  title={HOMER: a human organ-specific molecular electronic repository},
  author={Zhang, Fan and Chen, Jake Y},
  journal={BMC bioinformatics},
  volume={12},
  number={Suppl 10},
  pages={S4},
  year={2011},
  publisher={Springer}
}

@article{31,
  title={MEME SUITE: tools for motif discovery and searching},
  author={Bailey, Timothy L and Boden, Mikael and Buske, Fabian A and Frith, Martin and Grant, Charles E and Clementi, Luca and Ren, Jingyuan and Li, Wilfred W and Noble, William S},
  journal={Nucleic acids research},
  volume={37},
  number={suppl\_2},
  pages={W202--W208},
  year={2009},
  publisher={Oxford University Press}
}

@article{32,
  title={Deep learning for genomics: from early neural nets to modern large language models},
  author={Yue, Tianwei and Wang, Yuanxin and Zhang, Longxiang and Gu, Chunming and Xue, Haoru and Wang, Wenping and Lyu, Qi and Dun, Yujie},
  journal={International Journal of Molecular Sciences},
  volume={24},
  number={21},
  pages={15858},
  year={2023},
  publisher={MDPI}
}

@article{33,
  title={Deep Genomics: Deep Learning-Based Analysis of Genome-Sequenced Data for Identification of Gene Alterations},
  author={Kumar, Sourabh and Goel, Prabudh},
  journal={Artificial Intelligence (AI) in Cell and Genetic Engineering},
  pages={335--367},
  year={2025},
  publisher={Springer}
}

@article{34,
  title={Benchmarking interpretability of deep learning for predictive genomics: Recall, precision, and variability of feature attribution},
  author={Reynolds, Justin and Pan, Chongle},
  journal={PLOS Computational Biology},
  volume={21},
  number={12},
  pages={e1013784},
  year={2025},
  publisher={Public Library of Science San Francisco, CA USA}
}

@article{35,
  title={Interpretation of deep learning in genomics and epigenomics},
  author={Talukder, Amlan and Barham, Clayton and Li, Xiaoman and Hu, Haiyan},
  journal={Briefings in bioinformatics},
  volume={22},
  number={3},
  pages={bbaa177},
  year={2021},
  publisher={Oxford University Press}
}

@article{36,
  title={B-lineage commitment is dependent on a reversible epigenetic switch},
  author={Tingvall-Gustafsson, Johanna and Hellerstedt, Kim and Ungerb{\"a}ck, Jonas and Sigvardsson, Mikael},
  journal={Genes \& Development},
  volume={40},
  number={3-4},
  pages={233--254},
  year={2026},
  publisher={Cold Spring Harbor Lab}
}

@article{37,
  title={TAMC: A deep-learning approach to predict motif-centric transcriptional factor binding activity based on ATAC-seq profile},
  author={Yang, Tianqi and Henao, Ricardo},
  journal={PLOS Computational Biology},
  volume={18},
  number={9},
  pages={e1009921},
  year={2022},
  publisher={Public Library of Science San Francisco, CA USA}
}
\end{document}